\documentclass[a4paper,12pt]{article}

\usepackage[utf8x]{inputenc}
\usepackage{amsmath,amssymb}
\usepackage{graphicx}
\usepackage{caption}
\usepackage{cite}
\numberwithin{equation}{section}

\usepackage{lipsum}
\usepackage{lmodern}
\usepackage{tcolorbox}

\usepackage{empheq}
\usepackage{mathtools}

\graphicspath{ {images/} }
 \usepackage[toc,page]{appendix}

\usepackage{wrapfig}
\usepackage{float}
\usepackage{xcolor}
\usepackage{amsmath}
\usepackage{amsfonts}

\usepackage{hyperref}
\hypersetup{
    colorlinks=true,
    citecolor=blue,
    linkcolor=blue,
    filecolor=magenta,
    urlcolor=blue}

 \newcommand{\bea}{\begin{eqnarray}}
\newcommand{\eea}{\end{eqnarray}}

\newsavebox{\uuunit}
\sbox{\uuunit}
    {\setlength{\unitlength}{0.825em}
     \begin{picture}(0.6,0.7)
        \thinlines
        \put(0,0){\line(1,0){0.5}}
        \put(0.15,0){\line(0,1){0.7}}
        \put(0.35,0){\line(0,1){0.8}}
       \multiput(0.3,0.8)(-0.04,-0.02){10}{\rule{0.5pt}{0.5pt}}
     \end {picture}}

\newcommand{\lam}{\lambda}

\newcommand{\si}{\sigma}

\renewcommand{\cal}[1]{\mathcal{#1}}

\begin{document}
 

\title{\bf  Steady States of  Holographic Interfaces 
  \vskip 5mm
\,  }

\author{Constantin Bachas, Zhongwu Chen, Vassilis Papadopoulos }


\maketitle
\begin{center}
\textit{  Laboratoire de Physique  de l'\'Ecole Normale Sup\'eri{e}ure,\\
CNRS, PSL  Research University  and Sorbonne Universit\'es \\
24 rue Lhomond, 75005 Paris, France}
\end{center}

\vskip 1.5cm

\begin{abstract}

We find stationary thin-brane geometries that are dual to
far-from-equilibrium  steady states of  two-dimensional  holographic interfaces. 
The flow of heat at the  boundary agrees with   the result of CFT and the known   energy-transport
coefficients of the thin-brane  model. We argue that by entangling   outgoing excitations 
 the interface produces coarse-grained entropy at a  maximal  rate, and point out  similarities and differences with  double-sided
 black funnels. The
 non-compact,  non-Killing and  far-from-equilibrium event
  horizon  of our solutions  coincides with the local (apparent) horizon on the colder side,   but lies behind  it 
  on the hotter side of the interface. We also show that  the thermal  conductivity  of  a  pair
 of interfaces  jumps at  the Hawking-Page phase  transition  from a regime described by  classical  scatterers
 to a quantum regime in which heat flows  unobstructed.

    
\end{abstract}

\newpage

 {\small\tableofcontents}

\vskip 1.cm

\section{Introduction}

  AdS/CFT or holographic duality   
  \cite{Maldacena:1997re, Gubser:1998bc, Witten:1998qj}
    reduces  the study of certain strongly-coupled quantum systems  to (semi)classical equations in gravity. 
   The duality   has been mainly    tested  and exploited   at,  or   near  thermal equilibrium,  where  
   a hydrodynamic description applies.  For    far-from-equilibrium processes our  understanding is poorer.\footnote{ 
   See   \cite{Liu:2018crr} for  a recent review and references.} 
   Indeed,  although semiclassical gravity seems  more tractable,  highly-distorted  horizons raise a host of  
   unsolved technical and conceptual issues.
   To make progress, simple analytic models can be valuable. 
   We will study one such model here. 
   
   A  simple class of real-time processes are the non-equilibrium steady states (NESS)  
   characterised only by persistent currents. These are particularly simple  
      in critical  (1+1) dimensional
     ballistic systems   thanks to the power 
     of  conformal symmetry,  see \cite{Bernard:2016nci} for a review. 
     On the gravity  side the basic equilibrium states are the  Banados-Teitelboim-Zanelli (BTZ)
    black strings \cite{Banados:1992wn,Banados:1992gq}. 
     We deform the system by introducing a thin,   but strongly back-reacting,   domain wall  anchored 
    at   a conformal defect or interface on the  boundary.\footnote{  
     The wall   is a  Randall-Sundrum-Karch    brane \cite{Karch:2000gx,Bachas:2001vj}, but  
       crucially it is not an 
     End-of-the World brane:  Since we  are interested in heat flowing  across the
     interface,   there should be degrees of freedom  on both sides of  it. 
     } 
     Our goal is to compute   its  stationary states.

        Some  properties  of this  thin-brane holographic  model which   will be useful  later  have been    derived  recently
      in  
     refs.\cite{Bachas:2020yxv,Simidzija:2020ukv,Bachas:2021fqo}. We rederive in particular 
     the energy-transmission 
     coefficients obtained  from a scattering calculation  in \cite{Bachas:2020yxv}.  
 We also   revisit the Hawking-Page,  or  deconfinement  transition for  a theory contained between a
    pair   of interfaces  \cite{Simidzija:2020ukv,Bachas:2021fqo}, and show  that  at the critical temperature
     the thermal
     conductivity undergoes  
    a classical-to-quantum phase transition. 

    One  phenomenon  not discussed in these earlier works   is  the production of entropy. 
  This is due to scattering at the interface, which  entangles the   outgoing excitations thereby mixing
     the reflected and transmitted fluids. 
 A counter-intuitive feature of the thin-brane model  is that the interfaces  are 
 perfect scramblers --  the quantum  fluids exit,   as we will argue,   
  thermalised.\footnote{       A similar phenomenon is   the instant thermalisation of a holographic CFT  forced out of its 
     vacuum, as  discussed in \cite{Bhattacharyya:2009uu}.  We thank E. Kiritsis for pointing this out. 
     }
     Whether this feature survives in   
    top-down   solutions  with microscopic CFT duals is a  question left
     for future work.
    
        This  scrambling  behaviour is  reminiscent  of flowing black funnels
             \cite{Hubeny:2009ru,Fischetti:2012ps,Emparan:2013fha,Marolf:2013ioa,Santos:2020kmq},
             where a non-dynamical black hole acts as a
            source or sink of heat in the CFT.  There are however important differences between the two setups.  
           The  non-back-reacting 1+1 dimensional  black hole is a  spacetime boundary  that  can absorb 
           or emit arbitrary amounts of energy and entropy. Conformal interfaces, on the other hand, conserve energy
           and have a finite-dimensional Hilbert space. So even though one could mimic their
            energy and entropy flows  by a two-sided boundary  black hole 
           whose (disconnected)  horizon consists 
            of two points with  appropriately tuned  temperatures,  the rational, if any, behind such  tuning   is unclear. 
           
            The non-Killing event horizon of our solution is distorted far beyond the hydrodynamic regime.\footnote{
   For a review of the fluid/gravity correspondence see for instance \cite{Hubeny:2011hd}.}
   It  lies {\it behind}   the apparent  horizon which   is  the union of the BTZ horizons  on the two  sides  of the brane. 
   At the point where the brane enters the 
    event horizon this  latter  has discontinuous generators. Note that the event horizon is non-compact, thus evading theorems that   
    exclude  stationary non-Killing black holes
    \cite{Hawking:1971vc,Hollands:2006rj,Moncrief:2008mr}. The 
      apparent horizon is also not compact, both far from the brane and  at the brane-entry point. This prevents  a clash 
      with   
        theorems \cite{Hawking:1973uf} which show  that  event horizons  always  
         lie  {\it outside}  apparent (trapped-surface) ones.        
                      
           The plan of the  paper is as follows: In section \ref{sec:2n} we review 
           some well-known  facts about the BTZ 
           black string and   its holographic interpretation
           (savvy readers can skim rapidly through  this section). The black string is  a  rotating black hole  
          with   unwrapped angle variable and  spin $J$ equal to  twice  the  flow of  heat  in the dual CFT. For non-zero $J$ 
           there is  an ergoregion
          that plays a crucial role in our analysis.
                          
          In section \ref{sec:3n} we  explain, following   ref.\,\cite{Bernard:2014qia}, why  the   flow of heat 
          across  a 2d conformal interface   is proportional to the  energy-transmission coefficient(s) of the interface.  
          These  transport coefficients are universal,   independent of the nature  of   incident excitations  \cite{Meineri:2019ycm}. 
          By contrast,  the  coarse-grained entropy of the outgoing fluids  depends a priori  on  details 
           of the
          interface scattering matrix. If however  the  fluids are thermalised,   as     in our holographic model, 
           their  energy determines  their  (microcanonical)  entropy.

        Sections \ref{sec:4n} to  \ref{sec:7n} contain the 
          main results of our  paper. 
          In section \ref{sec:4n} and in appendix \ref{app:A} we solve  the  equations for a thin
          stationary   brane between  two arbitrary BTZ backgrounds. This generalizes
           the results of refs. \cite{Simidzija:2020ukv,Bachas:2021fqo} to  non-vanishing  BTZ spin $J$.  
           In section \ref{sec:5n} we show that  the brane penetrates the  ergoregion if and only if the heat flow 
           on the boundary agrees with the  
            prediction  from CFT  and with  the  transmission coefficients computed by a scattering calculation
            in  ref.\cite{Bachas:2020yxv}.  We also show that once inside
              the ergoregion the brane cannot exit towards  the AdS boundary, but crosses both  outer  BTZ horizons,  
              hitting eventually either a Cauchy horizon or the singularity in one of the two regions.

             Such a brane is  dual to  an  {\it isolated}  interface, and its
               non-Killing horizon   is computed 
          in section \ref{sec:6n}. We  show that it coincides with the  (local)  BTZ horizon on the colder side of the interface, and 
           lies behind but  approaches it asymptotically 
          on  the hotter side. This is the evidence   for  perfect scrambling mentioned above.        
           In section \ref{sec:7n} we consider  the system of  an 
            interface  pair which is known to have  an equilibrium  Hawking-Page 
            transition  \cite{Simidzija:2020ukv,Bachas:2021fqo}.  We show that thermal conductivity jumps discontinuously
            at the transition point, from a classical regime of stochastic scattering  to a deeply quantum regime
            in which   heat  flows unobstructed.            
            
              Section \ref{sec:8n} contains   closing remarks.  In order to not  interrupt    the flow of the paper
             we have relegated the proof of some  inequalities in 
             appendix 
            \ref{app:B},   and   background  material on flowing black funnels in appendix  \ref{app:C}  .


\section{The boosted AdS$_3$ black string}\label{sec:2n}
 
  The  boosted   black-string metric  of   three-dimensional  gravity with negative cosmological constant   reads
 \begin{equation}\label{metric}
ds^2  =     \frac{\ell^2 dr^2}{({r^2 } - M\ell^2  + {J^2 \ell^2/ 4r^2} )}   - ({r^2 } - M\ell^2 )dt^2  + r^2 dx^2  - J \ell  \,  dx dt \ ,  
\end{equation}
where  $x\in \mathbb{R} $ is non-compact. If  $x$ were an angle variable,  \eqref{metric}  
would be  the metric of the rotating BTZ black hole 
\cite{Banados:1992wn,Banados:1992gq} 
with   $M$  and $J$  its   mass and  spin\,\footnote{Strictly speaking these are defined with respect to
the rescaled time $t^\prime = t\ell$.  We will work throughout  in units $8\pi G=1$.}
 and $\ell$ the radius of AdS$_3$.

The   metric  \eqref{metric}   has an outer and an inner horizon   located at 
\bea\label{metric1}
r_\pm^2 = {\frac12} {M\ell^2 }   \pm {\frac12} \sqrt{ M^2\ell^4-J^2\ell^2}\, \,  \ . 
\eea
To avoid a naked singularity at $r=0$, one must require that $r_+$  be   real  which implies 
  $M \ell \geq \vert J \vert $.
In terms of $r_\pm$ the metric    reads
\begin{equation}\label{metric2}
\begin{split} 
&ds^2 =     {\ell^2 dr^2 \over h(r)}   - h(r) \, dt^2  +  (r \,dx  - {J \ell \over 2r}  \,  dt )^2\  , 
\\[1ex]
& \hskip -2cm {\rm with}\qquad  h(r) =  {1\over r^2} (r^2 - r_+^2)(r^2-r_-^2)  \qquad {\rm and} \qquad 
 \vert J \vert = {2r_+r_-\over \ell}   \ . 
\end{split}
\end{equation}
 Besides  $r_\pm$,   another special radius is  $r_{\rm ergo}  = \sqrt{M}\,  \ell \geq r_+ $. It  delimits
  the ergoregion inside which no observer (powered by any engine) can stay   at a  fixed
    position $x$.  





 Many  properties of the metric \eqref{metric2} are familiar from the Kerr black hole. See  \cite{Carlip:1995qv} for a nice review. 
 The outer horizon is a Killing horizon, while the 
  inner one  is a Cauchy horizon. Frame dragging forces    ingoing  matter  
   to  cross  the outer horizon  at  infinity along the string, $ x \sim J\ell \,t / 2r_+^2 \,\to\,\infty$. One can  define ingoing
 Eddington-Finkelstein (EF)  coordinates,   
\bea\label{EFc}
dv  =  dt  +   {\ell dr\over h(r)} \qquad {\rm and}\qquad  d y  = dx +  {J\ell ^2 dr   \over 2r^2  h(r)}  \ , 
\eea
in which  the metric 
\bea\label{EFmetric}
ds^2 \,=\,  - h(r)\, dv^{\,2} +   2\ell \, dv \, dr  + r^2 \big(dy  - {J\ell \over 2r^2}  \, dv \big)^2 \ 
\eea
is non-singular  at  the (future)   horizon.  
 Outgoing  coordinates  can be  defined similarly by changing $(x,t) \to (-x, -t)$
in \eqref{EFc}.


\subsection{Dual CFT$_2$   State}\label{sec:2.1n}
 
  In the context of AdS/CFT,  
  \eqref{metric}   describes  
 a non-equilibrium steady  state  (NESS) of  the CFT.  This has been  discussed in many places, see e.g. 
 \cite{Chang:2013gba,Bhaseen:2013ypa,Pourhasan:2015bsa,Erdmenger:2017gdk,Craps:2020ahu}.  
It  can be seen explicitly from 
 the    general asymptotically-AdS  solution of the vacuum  Einstein equations, whose 
  Fefferman-Graham expansion    in three dimensions terminates  \cite{Banados:1998gg}
  \begin{equation}\label{FG}
ds^2 = {\ell^2 dz^2 \over z^2 }  + {1\over z^2} \Bigl(dx^+ + \ell z^2  \langle T_{--} \rangle  dx^-\Bigr) 
\Bigl(dx^- + \ell z^2  \langle T_{++} \rangle   dx^+\Bigr)  \ .   
\end{equation}
Here $\langle T_{\pm\pm} \rangle$ are the  expectation values of the left-moving  and right-moving 
 energy densities in the dual CFT$_2$ state. 
 The  two  metrics,    \eqref{metric} and  \eqref{FG},   
can be  related  by   the  change of coordinates 
\begin{equation}
x^\pm = x\pm t \  , \qquad     r^2 =    {1\over z^2} \Bigl(1  + \ell z^2 \langle T_{--} \rangle  \Bigr) 
\Bigl(1  + \ell z^2  \langle T_{++} \rangle  \Bigr)  \  ,  
 \end{equation}
and  the   identification  
 \begin{equation}\label{J}
 \text{\footnotesize ${1\over 2}$} \,   J
 \,  = \,  \langle T_{--} \rangle - \langle T_{++}\rangle   \qquad {\rm and} \qquad 
  \text{\footnotesize ${1\over 2}$} \,   {M\ell} \, =\,   \langle T_{--}\rangle  +  \langle T_{++}\rangle  \ . 
\end{equation}
It follows that  the dual state has constant fluxes   of energy in both directions, with a net 
  flow    $ \langle T^{\,tx} \rangle =   J/2 $.  To abide with the standard notation for heat flow we will
  sometimes  write $J/2 = dQ/dt$.

 Generic  NESS  are  characterised by operators other than $T_{\alpha\beta}$, for instance by  persistent U(1)  currents. 
  To describe them  one must  switch  on non-trivial matter  fields,    and  the above 
  simple analysis must  be modified. The vacuum solutions 
  \eqref{metric} describe, nevertheless, a  universal class of NESS that exist in all holographic conformal theories.

   There are many  ways of preparing these universal  NESS. 
   One can   couple  the  endpoints  $x\sim \pm\infty$ 
  to heat baths  so that   left- and right-moving excitations  
  thermalise 
  at different   temperatures 
  $\Theta_\pm$\,.\footnote{We use $\Theta$ for temperature   to avoid
confusion with the energy-momentum tensor. In the gravitational dual the heat baths can 
be replaced by non-dynamical boundary black holes, see below. 
} 
 An alternative protocol (which avoids  the complications of  reservoirs and leads)  is  the  
  partitioning  protocol. Here 
  one   prepares   two semi-infinite systems at  temperatures $\Theta_\pm$, and   joins  them  
   at  some initial  time $t=0$. The steady state will  then  form  inside  a linearly-expanding interval   in the  middle
 \cite{Bernard:2016nci}. In both cases,  
after transients have died out  one expects 
    \begin{equation}\label{17}
\langle T_{\pm\pm} \rangle = {\pi c \over 12 }\, \Theta_\pm^2 =   {  \pi^2  \ell  }\,  \Theta_\pm^2\  \quad
\Longrightarrow \quad 
\langle T^{\,tx} \rangle = {\pi c\over 12}\, (\Theta_-^2  - \Theta_+^2) \  , 
\end{equation}
where   $c= 12\pi \ell$ is the   central charge of the CFT.
Equation \eqref{17} for the flow of heat
   is  a (generalized)  Stefan-Boltzmann law  with Stefan-Boltzmann constant $\pi  c/12$. Comparing
   \eqref{17}  to \eqref{J}  relates the temperatures $\Theta_\pm$ to the parameters $M$ and $J$ 
   of the black string.\footnote{ 
 This idealized CFT  calculation is,  of course,  only  relevant for systems  
in  which  the   transport of energy   is predominantly ballistic. Eq.\,\eqref{17} implies in particular  the existence
of a  quantum  of  thermal  conductance, see  the review \cite{Bernard:2016nci} and references therein. }
To  implement the partitioning protocol on the gravity side it is sufficient to  multiply  the  constant $\langle T_{\pm\pm} \rangle$
 in eq.\,\eqref{FG} by step functions $ \theta(\pm\, x^\pm)$.


             \begin{figure}[thb!]
   \vskip  - 2mm
\centering
\includegraphics[width=15cm]{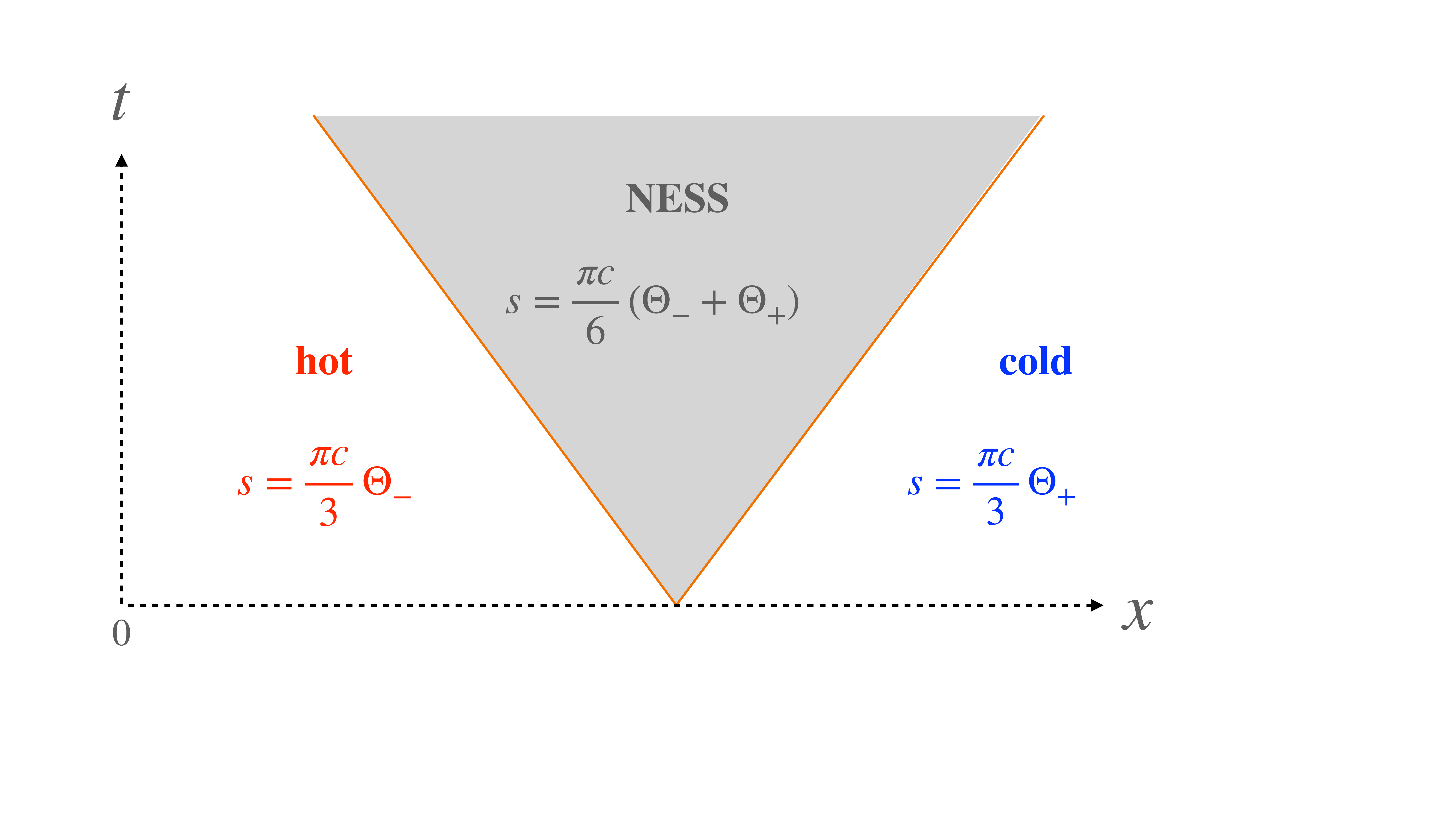}
\vskip -12mm
     \caption{\footnotesize When two identical semi-infinite quantum  wires  at temperatures $\Theta_\pm$ 
     are joined   at   $t=0$,  a  NESS  forms inside  an   interval that expands at constant speed  in both directions
       \cite{Bernard:2016nci}.  
     The entropy density $s(t,x) $  is shown  in the three regions of the protocol. 
   The  energy density  profile is identical, except  for   the replacement 
     $\Theta_\pm \to {1\over 2} \Theta_\pm^2$.
    }
\label{fig:protocols}
 \end{figure}

  It is  interesting  to also consider  the flow of entropy. This is illustrated   in 
  figure  \ref{fig:protocols} which 
shows the  entropy density  $ s \equiv \langle s^t \rangle$ in the three spacetime regions of   the partitioning   protocol. 
  Inside the NESS region there is constant flow of entropy from the hotter towards  the colder side
    \bea
\langle s_\pm \rangle  =  \pm {\pi c \over 6 }\, \Theta_\pm\ . 
\eea 
 The passage of the right-moving shock wave increases the local entropy at a rate $\pi c(\Theta_--\Theta_+)/6$,  
while the  left-moving wave reduces it at an equal  rate. Total entropy   is therefore  conserved,
not surprisingly  since there are no interactions  in this simple conformal 2D fluid. 


     
  One can compute  the entropy on the gravity side with the help of    
   the Hubeny-Rangamani-Ryu-Takayanagi  formula
    \cite{Ryu:2006bv,Hubeny:2007xt}. For  a boundary region of size  $\Delta x$  the entanglement entropy reads
  \cite{Hubeny:2007xt} 
\bea\label{qen}
 S_{q\,\rm ent} = {c\over 6} \log \big[ {\beta_+\beta_-\over \pi^2 \epsilon^2} \sinh ({\pi \Delta x\over \beta_+} ) \sinh ({\pi \Delta x\over \beta_-} )
 \big] \ , 
\eea
where $\beta_\pm = \Theta_\pm^{-1}$ and $\epsilon$ is a short-distance cutoff. 
From this one computes the  coarse-grained entropy density in  the steady state
  \bea\label{qen1}
  s_{\rm NESS}  =\ \lim_{\Delta x\to \infty}   { S_{q\,\rm ent}\over  \Delta x}  
  = {\pi c \over 6 }\, (\Theta_-+\Theta_+) = 2\pi\,  r_+ \ . 
  \eea 
The last equality,  obtained 
 with the help of   eqs.\,\eqref{17}, \eqref{J} and   \eqref{metric1},    recasts    $s_{\rm NESS} $
  as the Bekenstein-Hawking entropy of the boosted black string 
(recall that our  units  are $8\pi G = \hbar =1$).  This agreement was one of the earliest  tests
\cite{Strominger:1997eq}  of the AdS/CFT correspondence


  \section{NESS of   interfaces} \label{sec:3n}

Although  formally out-of-equilibrium, the  state of the previous section 
 is  a rather  trivial example of a NESS. It can be  obtained from
 the   thermal   state 
 by a Lorentz boost, 
and  is therefore a  
  Gibbs state  with   chemical potential  for  the  (conserved) momentum in the $x$ direction.

More interesting steady states  can be found when   left-  and right-moving  excitations  interact, for instance at 
     impurities \cite{Bernard:2014qia,Sonner:2017jcf,Novak:2018pnv}  
    or when the  CFT lives  in a non-trivial background  metric \cite{Hubeny:2009ru,Fischetti:2012ps,Marolf:2013ioa}. 
  Such    interactions lead to   long-range entanglement and decoherence,  giving   NESS  that are
 not just 
   thermal states in disguise.\footnote{\,Chiral separation also fails when  the CFT is deformed by  (ir)relevant     interactions.
    The  special case
   of the $T\bar T$   deformation  was studied,  using both  integrability and   holography,  in 
    refs.\cite{Medenjak:2020bpe,Medenjak:2020ppv}. Interestingly,   the  persistent energy current  takes again  the  
form \eqref{17} with a deformation-dependent  Stefan-Boltzmann  constant. }

\begin{figure}[thb!]
   \vskip  -5mm
\centering
\includegraphics[width=18cm]{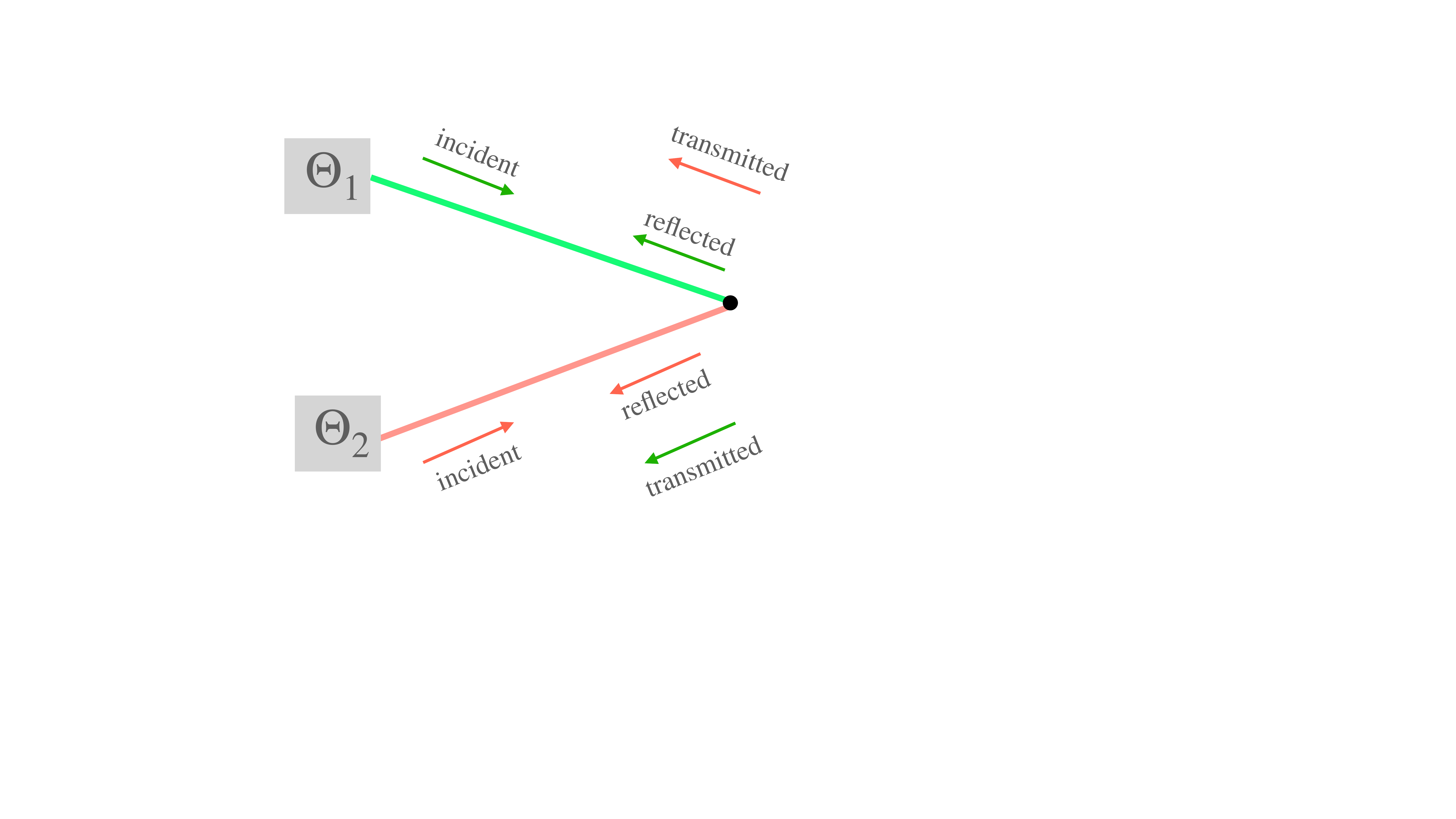}
\vskip -31mm
     \caption{\footnotesize The  energy fluxes given in \eqref{4flows}. The two half wires are coloured  red and green,
  and   space is  folded at the interface (black dot). The   incoming  excitations are thermal 
  while the state of the outgoing  ones, consisting  of both reflected and  transmitted fluids, depends on the nature of the junction as
discussed in the main  text. }
\label{fig:2}
 \end{figure}

 The case of  a conformal defect,  in particular,   has been  analyzed
   in ref.\cite{Bernard:2014qia}.  As explained   in this reference  
the heat current  is still given by  eq.\,\eqref{17} but the Stefan-Boltzmann  constant is multiplied by ${\cal T}$, 
   the energy-transmission coefficient of the defect. 
  The relevant setup is shown in figure \ref{fig:2}.  The fluids  entering the NESS region from opposite directions 
  are thermal    at  different  temperatures  
  $\Theta_{1} \not= \Theta_{2}$. The difference, compared to the discussion of the previous section, 
    is that the two half wires ($j=1,2$)  need not be identical, or   (even  when  they are)   their junction is a
  scattering impurity.

 

 \subsection{Energy currents}\label{sec:3.1n}

  If  ${\cal R}_{j}$ and ${\cal T}_{j}$ are  the reflection and transmission coefficients for energy  incident
 on the interface  from the $j$th side, then 
 the  energy  currents  in the   NESS read\,\footnote{\,The currents are given in  the folded picture in which the
 interface is a boundary of the tensor-product theory CFT$_1\otimes$CFFT$_2$, and both incoming waves depend on $x^-$.
 } 
 \begin{equation}\label{4flows} 
 \begin{split}
& \langle T_{--}^{(1)} \rangle =  { \pi c_1\over  12} \,\Theta_{1}^2 \ , \qquad
 \langle T_{++}^{(1)} \rangle =  {\cal R}_{1} { \pi c_1\over 12}\, \Theta_{1}^2  +  {\cal T}_{2} { \pi c_2\over  12}\, \Theta_{2}^2
  \ ,
 \\[1ex] 
  & \langle T_{--}^{(2)} \rangle =  { \pi c_2\over12 }\, \Theta_{2}^2 \ , \qquad
   \langle T_{++}^{(2)} \rangle =  {\cal T}_{1} { \pi c_1\over 12} \,\Theta_{1}^2  +  {\cal R}_{2} { \pi c_2\over 12} \,\Theta_{2}^2
  \ .  
  \end{split}
 \end{equation} 
We have used here  the key fact  that  the energy-transport  coefficients across  a 
  conformal interface in 2d  are universal, i.e independent  of  the 
 nature of the incident excitations.  The 
 proof  \cite{Meineri:2019ycm}
assumes that the  Virasoro symmetry is not extended by extra  spin-2 generators, 
which is true  in  our holographic model. We  have also used   that 
the incoming and outgoing excitations  do not  interact  away from  the interface.

  Conservation of energy and the detailed-balance condition (which ensures that when  $\Theta_{1}
 =\Theta_{2}$ 
 the heat flow stops) 
 imply the following relations among the reflection and transmission coefficients:
 \bea\label{detbal}
 {\cal R}_{j}+{\cal T}_{j} = 1\qquad {\rm and}  \qquad  c_1{\cal T}_{1}  =  c_2{\cal T}_{2}\ \,  . 
 \eea
Hence,  only one of the four transport  coefficients is  independent.
Without  loss of  generality we   assume that $c_2\geq  c_1$, i.e. that
  CFT$_2$  is the theory with more degrees of freedom.
  The average-null-energy  condition  requires    $0\leq {\cal R}_{j},  {\cal T}_{j} \leq 1$, so from  \eqref{detbal} 
  we conclude 
   \bea
 0\leq {\cal T}_{2} \leq  {c_1\over c_2}  \qquad  {\rm or \ equivalently}    \qquad 1 \geq {\cal R}_{2}  \geq    1- { c_1\over c_2}\,\ .  
 \eea
As noticed in
 \cite{Meineri:2019ycm},   reflection positivity of  the Euclidean theory gives   a weaker bound 
  \cite{Billo:2016cpy} than this  Lorentzian bound. Note also that in 
  the asymmetric case  ($c_2$ strictly bigger than $c_1$)   part  of the energy incident   from side 2  is  necessarily
 reflected.

Let $dQ/dt =  \langle T^{(1)\,tx}  \rangle  = - \langle T^{(2)\,tx}\rangle $ be the heat current across the interface.
From eqs.\,\eqref{4flows} and \eqref{detbal} we find
   \bea\label{BD1}
  {dQ \over dt}  =  {\pi\over 12} c_1{\cal T}_{1} \big(
   {  \Theta_{1}^2 } -  {  \Theta_{2}^2 } \big) \ . 
  \eea
Since in a unitary theory 
  $c_1 {\cal T}_{1} $ is non-negative,    
    heat    flows  as expected from the hotter to the colder side. The heat flow  only stops    for 
  perfectly-reflecting interfaces  (${\cal T}_{1} = {\cal T}_{2}= 0$),  or when  the two  baths are at equal  temperatures. 
  For small  temperature difference, the heat conductance reads
 \bea
  {d  Q \over dt } \, =\,  {\pi \Theta  \over 6} c_j{\cal T}_{j} \,   \delta\Theta   \ . 
 \eea
The  conductance per degree of freedom, ${\pi \Theta / 6}$, is thus
multiplied by the transmission coefficient of the defect \cite{Bernard:2014qia}.
     Note finally   that the
  interface is   subject to  a radiation force 
given by   the   discontinuity of pressure, 
 \bea
 F_{\rm rad} =  \langle T^{(1)\,xx}\rangle  -  \langle T^{(2)\,xx}  \rangle\  = \  {\pi\over 6} \big( {c_1 {\cal R}_1   \Theta_1^2} - 
 {c_2 {\cal R}_2  \Theta_2^2}
 \big) \  , 
\eea
where we   used   eqs.\,\eqref{4flows} and \eqref{detbal}. 
The  force   is proportional to the reflection coefficients, 
 as expected.

 
 \subsection{Entropy production}\label{sec:3.2n} 
 
    There is a crucial  difference between the NESS of  section \ref{sec:2.1n}, and the NESS  in the presence of the interface. 
   In both cases the incoming  fluids are  in a thermal state. But while for  
   a  homogeneous wire  they  exit the system  intact, in the presence  of an interface they interact and become   entangled. 
   The  state of the outgoing  excitations  depends therefore  on the nature of these interface   interactions.

\begin{figure}[thb!]
  \vskip  -3mm
\centering
\includegraphics[width=14.5cm]{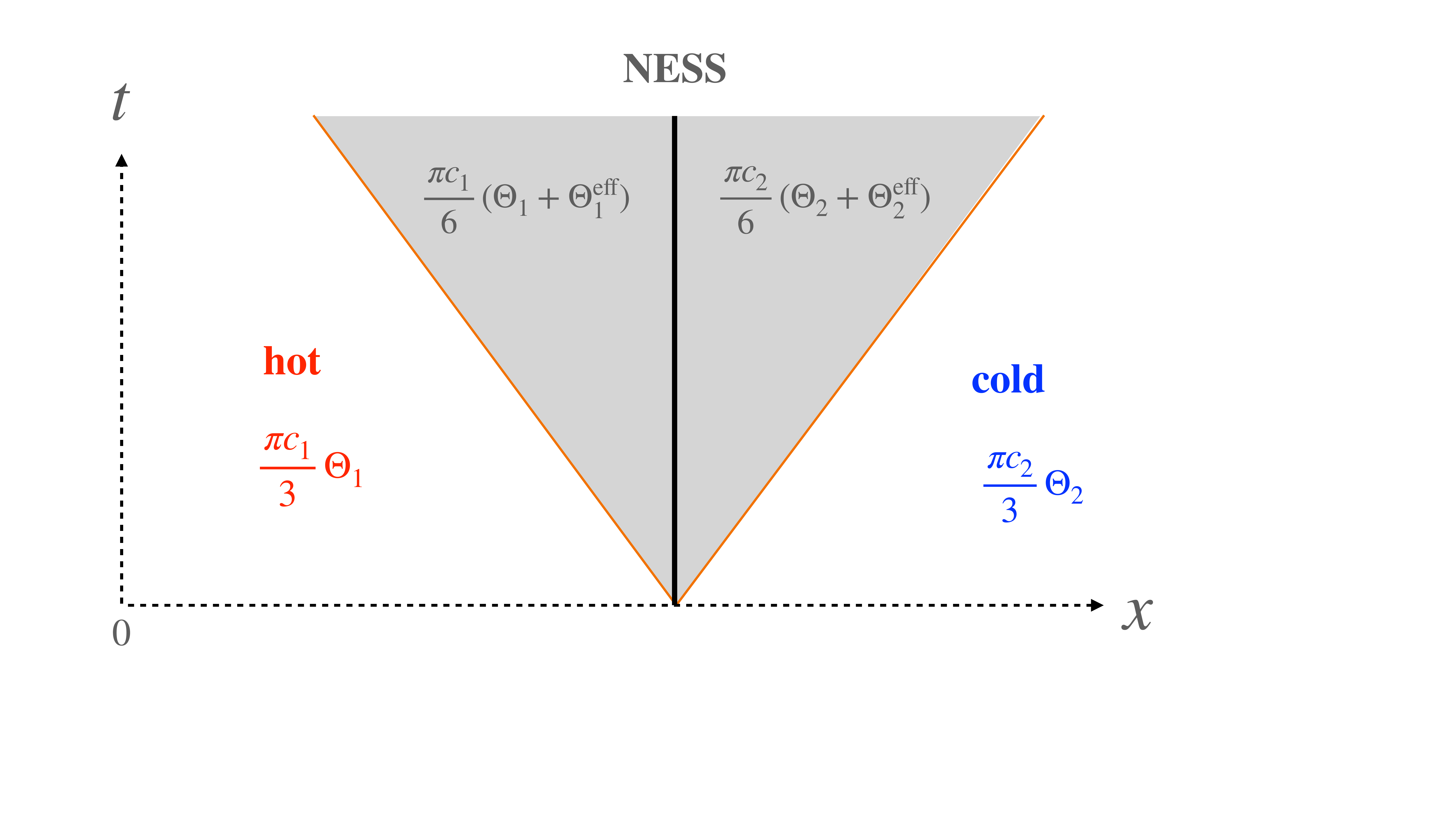}
\vskip -14mm
     \caption{\footnotesize   The entropy densities in the four regions of the partitioning protocol 
     discussed in the text (space  is here unfolded).  
     The entropies of the outgoing fluids, which depend a priori on details of the scatterer, have been parametrized
  by two  effective temperatures. 
   }
\label{fig:protocols2}
 \end{figure}

  Let us  consider the coarse-grained  entropy density  of the outgoing fluids, defined as 
 the von Neumann entropy density  for an interval   $[x, x+ \Delta x]$.  
   We parametrise it 
    by  effective temperatures, so that the entropy currents read
   \begin{equation}\label{4S} 
 \begin{split}
& \langle s_{-}^{(1)} \rangle =  -{ \pi c_1\over  6} \,\Theta_{1}  \ , \qquad
 \langle s_{+}^{(1)} \rangle =  { \pi c_1\over  6} \,\Theta_{1}^{{\rm eff}}
  \ ,
 \\[1ex] 
  & \langle s_{-}^{(2)} \rangle =  -{ \pi c_2\over 6 }\, \Theta_{2}  \ , \qquad
   \langle s_{+}^{(2)} \rangle =  { \pi c_2\over  6} \,\Theta_{2}^{{\rm eff}}
  \ .  
  \end{split}
 \end{equation} 
 We stress   that  \eqref{4S}  is just  a parametrisation,   the outgoing fluids need not be   in a thermal state. 
  In principle $\Theta_j^{\rm eff}$ may vary  as a function of $x$, but we expect them to  approach
 constant values   in the limit   $t\gg \vert x\vert \gg \Delta x \to  \infty$. 
 Figure \ref{fig:protocols2} is a cartoon of  the 
  entropy-density profile $\langle s^t\rangle$  in various spacetime regions of the partitioning protocol. 
Entanglement  at the interface produces thermodynamic  entropy that  
is  carried away by the two shock waves at a rate      
   \begin{equation}\label{4Sflows} 
    {dS_{\rm tot} \over dt} = {\pi c_1\over 6}  (\Theta_1^{\rm eff}  - \Theta_1) +   {\pi c_2\over 6}  (\Theta_2^{\rm eff}  - \Theta_2)
    +   {dS_{\rm def}\over dt} \  
 \end{equation} 
 where $S_{\rm def}$ denotes  the entropy of the interface.  Since this  is bounded by the logarithm of the   $g$-factor, 
  $S_{\rm def}$ cannot grow indefinitely and 
  the last  term of  \eqref{4Sflows}   can be neglected  in a   steady state.\,\footnote{ 
Defects  with  an infinite-dimensional Hilbert space  may evade this argument.
  But in the holographic model studied in this paper,   $\log g \sim O(c_j)$
   \cite{Azeyanagi:2007qj,Simidzija:2020ukv} and the last term in 
   \eqref{4Sflows} can be again   neglected at   leading semiclassical order.
 }

 The entanglement between outgoing excitations is encoded in a 
   scattering matrix, which we may write schematically as
       \bea\label{Smatrix}
   \mathbb{S}(\psi_1^{\rm in}, \psi_2^{\rm in}, \psi_{\rm def}^{\rm in}\,\vert \,  \psi_1^{\rm out},  \psi_2^{\rm out}, \psi_{\rm def}^{\rm out}  )\ .  
 \eea
Here $\psi_j^{\rm in/out}$ are the  incoming  and outgoing excitations,  and $\psi_{\rm def}^{\rm in/out}$  is the  state of the defect before/after the scattering.  Strictly speaking  there is no  genuine  S-matrix in  conformal field theory. 
What describes the conformal interface is
a formal operator ${\cal I}$, obtained by unfolding the associated boundary state  \cite{Bachas:2001vj,Bachas:2007td}. 
   The above  $\mathbb{S}$ is an  appropriate Wick rotation of ${\cal I}$, as explained   
   in ref.\cite{Bernard:2014qia}.\,\footnote{ The (closed-string channel) operator ${\cal I}$ evolves the system  across  a quench,
   whereas $\mathbb{S}$ should be  defined in real time in the open-string channel. 
A   careful discussion of   `collider experiments' in CFT$_2$ is also given in ref.\cite{Meineri:2019ycm}.}  
The density matrix  of the outgoing fluids depends a priori   on the entire  S-matrix,  not just on the  transport
 coefficients ${\cal T}_j$ and ${\cal R}_j$. 

   The second law of thermodynamics bounds the effective temperatures from below 
   since the entropy production 
\eqref{4Sflows}  cannot  be    negative. The  $\Theta_j^{\rm eff}$ are  also   bounded from above because  the entropy
density cannot exceed the  microcanonical  one,    $ s =  (\pi c\, u/3)^{1/2}$ 
  with  $u$ the energy density of the chiral fluid. Using   \eqref{4flows} and  
  the  detailed-balance condition this gives 
 \bea\label{boundsS}
 \Theta_{1}^{{\rm eff}} \leq \sqrt{ {\cal R}_1 \Theta_1^2 + {\cal T}_1 \Theta_2^2 } 
 \qquad {\rm and} \qquad
  \Theta_{2}^{{\rm eff}} \leq \sqrt{ {\cal R}_2 \Theta_2^2 + {\cal T}_2 \Theta_1^2 } \ \ . 
\eea
The   bounds  are saturated   by perfectly-reflecting or transmitting interfaces,  i.e. when  
either ${\cal R}_j=1$ or  ${\cal T}_j=1$.  This is trivial, because in such cases  
 there is no   entanglement between the outgoing fluids. 

 Partially reflecting/transmitting interfaces that saturate the bounds \eqref{boundsS}  act as   
perfect scramblers. 
 Their existence  at weak coupling seems  unlikely,  but   strongly-coupled holographic interfaces could  be of this kind.
 We will later argue  that the  thin-brane holographic interfaces are perfect scramblers. This 
    is supported by the fact (shown   in section \ref{sec:62n})
  that  far from the brane the event horizon  approaches the equilibrium BTZ horizons, and 
hence  the outgoing  chiral fluids are thermalised.   
 
  Any domain-wall solution  interpolating 
    between  two BTZ geometries,   with  no other  non-trivial  asymptotic 
   backgrounds  should  be likewise  dual to a  NESS of a  perfectly\,-\,scrambling  interface. We  suspect that  
   many top down  solutions  of this kind  exist, but they are hard to find. 
  Indeed,  although many  BPS  domain walls  are known   in  the supergravity  literature,   
 their  finite-temperature counterparts  are  rare. 
   The one  example that  we are aware of   is the Janus AdS$_3$ black brane  
     \cite{Bak:2011ga}. But  even for  this  computationally-friendly   example 
   the     far-from-equilibrium stationary  solutions  are not known.


 
\section{Stationary branes}\label{sec:4n} 
   
        To simplify the problem   we will here  resort  to
     the more tractable   thin-brane approximation, hoping that it  captures some of  the essential physics of the stationary  states. 
     This  thin-brane  holographic model  is also  the one   studied   in the related  
      papers \cite{Bachas:2020yxv,Simidzija:2020ukv,Bachas:2021fqo}. 
     
 
 \subsection{General setup}

   Consider   two  BTZ metrics   \eqref{metric}  glued  along a thin  brane 
whose  worldvolume  is parametrised 
   by  $\tau$ and  $\sigma$. Its 
 embedding    in the two coordinate patches  ($j=1,2$)  is   given by
 six functions $\{ r_j(\tau, \sigma), \,t_j(\tau, \sigma), \,x_j(\tau, \sigma)\}$.
       The most general  stationary   ansatz, such 
that   the induced   metric   is $\tau$-independent,  is of the form
\begin{equation}\label{ansatz} 
 x_j(\sigma), \ \ r_j(\sigma), \ \ t_j =      \tau + f_j(\sigma)\ . 
\end{equation} 
In principle one  can multiply $\tau$ on the right-hand side by  constants $a_j^{-1}$. 
  But   the metric \eqref{metric} is invariant under   rescaling  of the coordinates $r \to a r $,  $(t ,  x) \to a^{-1} (t, x) $,
 and of the parameters  $(M,  J) \to a^{2} (M, J) $, so we  may  absorb the $a_j$ into  a redefinition of  the parameters
 $M_j, J_j$. Hence,    without loss of generality,  we set
 $a_j = 1$.      
 
    Following ref.\cite{Bachas:2021fqo}  we  choose  the parameter $\sigma$  to be the   
    redshift\ factor squared\,\footnote{This is a slight misnomer, since $\sigma$ becomes  negative
  in  the ergoregion.} for  a stationary observer
   \bea\label{rj1}  
\sigma \, =\,  r_1^2 -M_1\ell_1^2 \, =\,  r_2^2 -M_2\ell_2^2  \ . 
\eea 
With this choice   $\hat g_{\tau\tau}=-\sigma$\,  is the same on the two sides of  the domain wall, and  
 the functions    $r_j(\sigma)$ are determined. 
   Of  the remaining embedding  functions,  the sum $f_1+f_2$   is pure gauge (it can be absorbed by a
   redefinition of $\tau$) whereas
  the time  delay across  the   wall, 
     $\Delta t (\sigma) \equiv f_2(\sigma) - f_1(\sigma)$,    is a physically quantity. This and the two  functions $x_j(\sigma)$  
   should be determined  by solving the  three remaining  
equations: (i)\,the continuity of the induced-metric components $\hat g_{\tau\sigma}$ and $\hat g_{\sigma\sigma}$, and 
(ii) one of the
(trace-reversed) 
 Israel-Lanczos  conditions\,\footnote{Two 
 of the three  Israel   conditions are automatically satisfied, modulo  integration constants,
by virtue of  the momentum constraints $D^\alpha K_{\alpha\beta} - D_\beta K=0$.
Our conventions are the same as in  ref.\cite{Bachas:2021fqo}:  
$K_{\alpha\beta}$ is the covariant derivative of the inward-pointing unit normal vector,
and the orientation is chosen so that 
as  $\sigma$ increases  one encircles  {\it clockwise} the interior  in the $(x_j, r_j^{-1})$ plane,  
in both charts. }
 \bea
 \label{IsLan}
 [K_{\alpha\beta}]  =  - \lambda \,  \hat g_{\alpha\beta} \ . 
 \eea
Here $K_{\alpha\beta}$ is the extrinsic curvature (with $\alpha, \beta \in \{ \tau, \sigma\}$), the brackets denote
the discontinuity across the wall,  and 
$\lambda$ is the brane tension. 
 

 \subsection{Solution of  the  equations}\label{sec:4.2n}

The general  local solution of the  matching equations is derived  in   appendix \ref{app:A}. 
 The solution is given   in the `folded setup' where  the interface is a conformal boundary for  the
  product theory CFT$_1\otimes\,$CFT$_2$. Unfolding side $j$ amounts to sending $x_j \to -x_j$ and $J_j\to -J_j$.

   The results of appendix \ref{app:A}  can be  summarised  as follows. First,   from \eqref{rj1} 
    \bea\label{rj}
  r_j(\sigma)    = \sqrt{\sigma + M_j \ell_j^{\,2}}\ .  
  \eea
  Secondly, consistency of the extrinsic-curvature equations  imposes 
  \bea
  J_1 = -J_2  \ . 
  \eea
This ensures  conservation of energy in the CFT, as seen from  the holographic 
 dictionary \eqref{J}. Thirdly,   matching $\hat g_{\tau\sigma}$ from  the two sides determines  the 
   time delay   in terms of the embedding 
   functions $x_j$, 
 \bea\label{fj} 
\Delta t^\prime  \equiv f_2' - f_1'   =     { J_1 \over 2 \sigma } ( \ell_1 x_1^\prime + \ell_2 x_2^\prime)\ ,
\eea
where primes denote derivatives with respect to $\sigma$. 
What remains  is thus to find  the functions  $x_j(\sigma)$.

To this end we use the  continuity
 of  $\hat g_{\sigma\sigma}$ and  the $\tau\tau$ component of  \eqref{IsLan}.
   It is  useful and convenient to first solve these two equations for the determinant of the induced metric, with the result
\bea\label{detg}
 -{\rm det}\, \hat g \,=\,    {\lambda^2 \sigma \over  {A \sigma^2 + 2B\sigma +C }} \, =\, 
  {\lambda^2 \sigma \over  {A (\sigma - \sigma_+)(\sigma - \sigma_-)}}\ ,  
\eea  
where 
 \bea\label{spm} 
 \sigma_\pm =  {-B \pm \sqrt{B^2 - AC} \over A} 
 \eea 
 and the coefficients $A,B,C$ read
  \begin{equation}\label{coeff}
\begin{split} 
   A = (\lambda_{\max}^2-\lambda^2) (\lambda^2-\lambda_{\min}^2), & \quad 
B = \lambda^2(M_1+M_2) - \lambda_0^2(M_1-M_2), 
\\[1ex]
&  \hskip -1cm   C = -(M_1-M_2)^2 + \lambda^2J_1^2\ .   
\end{split}
\end{equation} 
 The three critical tensions entering  in the above coefficients have been   defined 
 previously in refs.\cite{Bachas:2020yxv,Bachas:2021fqo}, 
   \bea\label{3crtens}
  \lambda_{\rm min}  =   \biggl\vert {1\over \ell_1} - {1\over \ell_2} \biggl\vert  \ , \quad
 \lambda_{\rm max} = {1\over \ell_1}+  {1\over \ell_2}\ , \quad
 \lambda_0  = \sqrt{\lambda_{\rm max}\lambda_{\rm min}}\ .   
\eea
 Without loss of generality we  assume,  as earlier, that   $\ell_1\leq \ell_2$,  so 
 the absolute value in   $ \lambda_{\rm min} $ is superfluous. Note that the expressions \eqref{detg}  to \eqref{coeff} 
 are the same as the ones for  static branes   \cite{Bachas:2021fqo} except  for the extra term 
 $\lambda^2 J_1^2$ in the  coefficient  $C$.

  The  determinant  of  the induced metric  can be expressed in terms of $x_j$ and $\sigma$ in each chart,   $j=1$ and $j=2$. 
  It does not depend on the   time-shift functions  $f_j$, which could  be absorbed
  by a reparametrisation of the metric with  unit Jacobian. Having  already extracted  ${\rm det}\, \hat g$, one  can
  now invert these relations to find the $x_j'$, 
    \begin{equation}\label{xj11} 
    \frac{x_1'}{\ell_1} = -  \frac{{\rm sgn}(\sigma) \big[ (\lambda^2+\lambda_0^2)\,\sigma^2 \, + (M_1-M_2)\sigma \big] }
     {2(\sigma- \sigma_+^{\rm H1} )(\sigma- \sigma_-^{\rm H1}) 
    \sqrt{A\sigma  (\sigma - \sigma_+)(\sigma - \sigma_-) }}\ ,  
\end{equation}
\begin{equation}\label{xj22} 
  \frac{x_2'}{\ell_2} = -  \frac{{\rm sgn}(\sigma) \big[(\lambda^2-\lambda_0^2)\,\sigma^2 \, - (M_1-M_2) \sigma \big] }
   {2(\sigma- \sigma_{+}^{\rm H2} )(\sigma- \sigma_-^{\rm H2}) 
    \sqrt{A \sigma  (\sigma - \sigma_+)(\sigma - \sigma_-) }}\ , 
\end{equation}  
where here
\bea\label{sH}
\sigma_{\pm}^{{\rm H}j} =  - {M_j\ell_j^2\over 2}   \pm  {1 \over 2}\sqrt{M_j^2 \ell_j^4 -J_j^2 \ell_j^2 } \  
\eea  
are the  points where   the outer and inner horizons of  the $j$th  BTZ metric  intersect the domain wall. 
 
 Eqs. \eqref{rj} to  \eqref{sH} give  the general stationary  solution of the thin-brane  equations  for
 any Lagrangian parameters $ \ell_j$ and $\lambda$, and  geometric parameters
  $M_j$ and $J_1=-J_2$.  The Lagrangian parameters are part of the basic data of the interface CFT, 
  while the geometric parameters determine the CFT  state. 
 When  $J_1=J_2=0$,  all  these expressions  reduce  to the   static  solutions   found
  in ref.\cite{Bachas:2021fqo}.

 
 \section{Inside  the ergoregion}\label{sec:5n} 
 
  The qualitative behaviour of the domain wall is governed  by  the  singularities of  (\ref{xj11}, \ref{xj22}),  as one moves from
  the AdS boundary  at $\sigma \sim  \infty$     inwards. 
 In addition to the BTZ horizons at $\sigma^{{\rm H}j}_\pm$, other potential singularities arise  at $\sigma_\pm$
 and  at the entrance  of the ergoregion $\sigma=0$.  
From \eqref{detg}  we see  that the brane worldvolume would become  spacelike beyond    $\sigma=0$,
if $\sigma_\pm$ are both  either  negative or complex. 
  To avoid such pathological behaviour  
 one of the following two conditions must  be met:
 
 \begin{itemize}

\item    \fbox{$\,\sigma_+>0 \,:$}  The singularity at $\sigma_+$  is in this case   a turning
point, and the   \\  \vskip -6.2mm wall does not extend  to lower values of $\sigma$. Indeed, as seen from    \eqref{xj11} and \eqref{xj22},  
$dr_j/dx_j \vert_{\sigma_+} =0$  
 and the  $\sigma_+$  singularity is integrable, 
i.e.  the wall  turns around  at finite   $x_j = \int x_j'$\,. 
 
\item   \fbox{$\,0= \sigma_+  > \sigma_-\,:$} In this case the  worldvolume remains timelike as the wall  
 enters  the ergoregion. The reader can verify from 
  eqs.\,\eqref{detg}, \eqref{xj11} and  \eqref{xj22} that the embedding near  $\sigma=0$ is smooth.

\end{itemize}

           \begin{figure}[thb!]
   \vskip  -1mm
\centering
\includegraphics[width=1.03\textwidth,scale=0.85,clip=true]{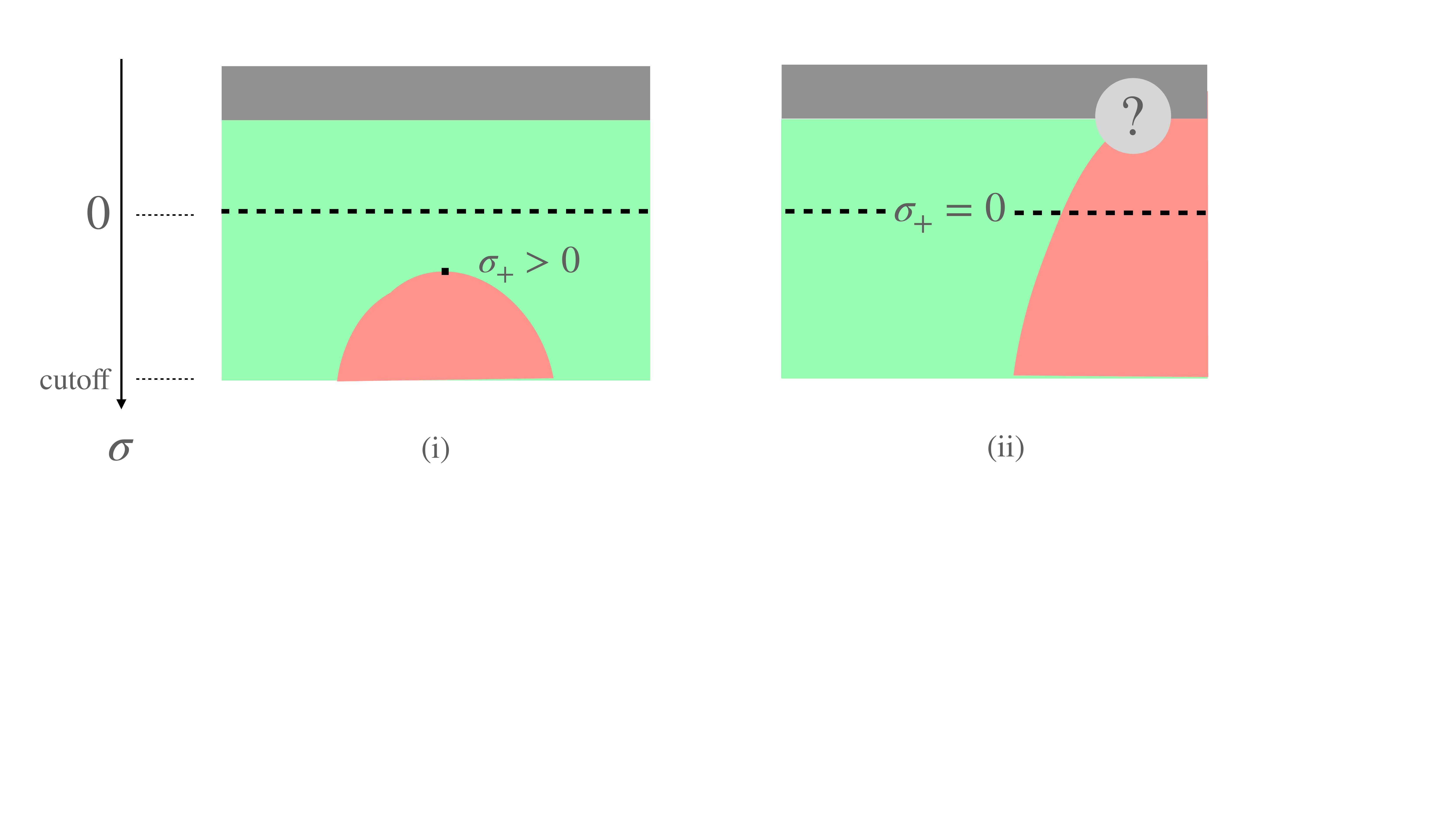}
\vskip -33mm
     \caption{\footnotesize The  two kinds of   stationary-wall geometries:  (i) 
      The wall avoids  the ergoregion, turns around   and intersects  the AdS boundary twice; or (ii) it  enters the ergoregion
      and  
     does not come out again.  
 The broken line is the ergoplane, 
 the two outer BTZ
 regions  are coloured in  green and pink, and the region behind the horizon in  grey. 
 The    horizon  in case (ii) will be described in detail  in the coming section.  }
\label{fig:4}
 \end{figure}    
     
 These  two   possibilities  are illustrated    in 
   figure \ref{fig:4}.  
    Branes   entering the ergoregion  are dual, as will become clear,    to steady  states  of an isolated 
   interface, while those that avoid  the ergoregion  are dual to  steady states  of  an interface-antiinterface pair.
We will return to the second  case in section \ref{sec:7n}, here we focus on the isolated  interface.

   The condition  $\sigma_+=0$  implies  $C=0$ and $B\geq 0$. Using  eqs.\,\eqref{spm} and \eqref{coeff},  and the fact
   that the coefficient   $A$ is positive  
for  tensions   in the allowed range
  ($\lambda_{\min} < \lambda < \lambda_{\max}$) we obtain
  \bea\label{s+0}
    M_1-M_2 =  \pm \lambda  J_1 = \mp \lambda J_2  
\quad
  {\rm and} \quad 
   \lambda^2 (M_1+M_2)  \geq  \lambda_0^{\,2} (M_1-M_2) \ . 
  \eea
Furthermore,  cosmic censorship requires that   $\ell_j M_j> \vert J_j\vert$
 unless 
 the bulk  singularity at  $r_j=0$  is  excised  (this is the case  in  the pink region  of the left  figure\,\ref{fig:4}).  
If none of the singularities is excised,   the inequality in \eqref{s+0} 
is  automatically satisfied and hence redundant.

With the help of 
  the  holographic dictionary  \eqref{J} one  can  
  translate the expression  \eqref{s+0} for $M_1-M_2$  to the   language of ICFT. 
Since the  incoming fluxes are thermal,\,   $T_{--}^{(j)} = \pi^2 \ell_j \Theta_{j}^2$\, and \eqref{J}  gives
        \bea
    M_j =   {4\pi^2 \Theta_{j}^2 }  - {J_j\over \ell_j}\  \ \Longrightarrow \  \ 
    M_1 - M_2 =  4\pi^2 ( \Theta_{1}^2  -  \Theta_{2}^2 )   - J_1 ({1\over \ell_1} + {1\over \ell_2}) \ .  
    \eea
Combining with eq.\,\eqref{s+0}   gives the  heat-flow rate
 \bea
 J_1 \,=\,  2  \langle T^{(1)\,tx}  \rangle \ = \    {4\pi^2} \big[  {1\over \ell_1} + {1\over \ell_2} \pm \lambda \big]^{-1} 
( \Theta_{1}^2  -  \Theta_{2}^2 ) \ .  
 \eea
This agrees with the ICFT  expression  \eqref{BD1} if we identify the transmission coefficients  as follows
  (recall that $c_j=12\pi \ell_j$)
 \bea\label{Tj1}
{\cal T_j} = {2\over \ell_j}\big[ {1\over \ell_1} + {1\over \ell_2}  \pm \lambda \big]^{-1} \ . 
\eea
It is gratifying to find   that,  for the choice of  plus sign, 
 \eqref{Tj1}  are precisely   the coefficients ${\cal T_j} $ computed in the  linearized 
 approximation
in ref.\,\cite{Bachas:2020yxv}. 
The  choice  of sign will be justified in a minute. 

 Let us pause here to take stock of the situation. We found that  (i) the dual of an isolated   interface {\it must} 
 correspond  to a  brane that enters  the ergoregion, and (ii) that  the brane equations determine in this case the flow
 of heat as expected  from  the CFT  result of  \cite{Bernard:2014qia} and the transmission coefficients derived in 
\cite{Bachas:2020yxv}. To complete the story, we must make sure  that once inside  the ergoregion the brane does not
come out  again. If it did,   it would intersect the AdS boundary at a second point, so the solution would not
be  dual to an isolated interface as claimed.

Inserting    $\sigma_+=0$ in  the embedding functions    (\ref{xj11},\,\ref{xj22}) we find 
 \begin{equation}\label{xj} 
    \begin{split}
 &   \frac{x_1'}{\ell_1} = -  \frac{(\lambda^2+\lambda_0^2)\,\sigma  \, + (M_1-M_2)} {2(\sigma- \sigma_+^{\rm H1} )(\sigma- \sigma_-^{\rm H1}) 
    \sqrt{A (\sigma - \sigma_-) }}\ ,  
\\[1ex]
&
  \frac{x_2'}{\ell_2} = -  \frac{(\lambda^2-\lambda_0^2)\,\sigma  \, - (M_1-M_2) }
   {2(\sigma- \sigma_+^{\rm H2} )(\sigma- \sigma_-^{\rm H2}) 
    \sqrt{A  (\sigma - \sigma_-) }}\  ,  
    \end{split}
\end{equation}  
where  the   $\sigma_\pm^{{\rm H}j}$ are given by  eq.\,\eqref{sH} and   
 \bea\label{s--}
\sigma_- = -\,  {2\lambda  \over A} \big[ 
  \lambda (M_1+M_2) \pm  2  \lambda_0^2J_1 \big]   \ .
\eea
As already said, the  embedding    
is regular   at $\sigma=0$, i.e. the brane  enters the ergoregion smoothly.  What it does  next depends on which 
singularity it encounters first. If this were the square-root singularity at 
  $\sigma_-$\,,  the  wall 
    would    turn around (just like it does for  positive $\sigma_+$),  
  exit the ergoregion and intersect  the AdS boundary at another  anchor point.   
 This is the possibility that we want to exclude. 

 Consider for starters  the simpler case  $\ell_1 = \ell_2 \equiv \ell$. 
 In this  case $\lambda_0=0$ and $A = \lambda^2 ( 4/\ell^2 - \lambda^2)$, so  \eqref{s--}  reduces to  
 \bea\label{ineq22}
  \sigma_-  \,= \,   -  {2\ell^2 \, (M_1+ M_2)  \over 4- \lambda^2\ell^2} \  \leq     \  - {\rm min}(M_j)\, \ell^2 \ . 
  \eea 
In the last step we   used the fact that  both $M_j$ are positive, otherwise the conical  singularity  at $r_j=0  \Longleftrightarrow
\sigma =  - M_j \ell^2$ would be naked. 
What \eqref{ineq22}  shows 
  is that   the putative 
turning point $\sigma_-$  lies behind the bulk  singularity in at least one of the two
BTZ regions,  where our solution cannot be extended. Thus this turning point is never reached.

   For general $\ell_1 \not=\ell_2$ a weaker  statement is   true, namely that  $\sigma_-$ is   shielded by  
  an inner  horizon for  at least one $j$. 
     The  proof requires   maximising   $\sigma_-$ with respect to the brane tension $\lambda$. 
  We have performed this  
  calculation  with Mathematica, but   do not find it  useful to reproduce   the nitty gritty  details here. The key point 
  for our purposes  is that there are no  solutions in which the  brane enters the ergoregion, turns around before an inner horizon, 
   and exits towards the AdS boundary. Since 
    as argued by Penrose  
 \cite{penrose1974gravitational},  Cauchy (inner) horizons are   classically 
 unstable,\footnote{For recent  discussions  of  strong cosmic censorship in the
 BTZ black hole see  \cite{Dias:2019ery,Papadodimas:2019msp,Balasubramanian:2019qwk,Emparan:2020rnp,Pandya:2020ejc}.}
solutions in which the turning point lies behind one of them cannot be trusted.

One last remark is in order concerning  the induced  brane metric $\hat g_{\alpha\beta}$. 
By redefining   the worldvolume time,   $\tilde{\tau}=\tau+ {J\ell_1} \int  {x_1'(\si)}d\si /{2\si}$,   we can  bring   
this metric   to the diagonal form 
\begin{equation}
    d\hat s^2 =-\si d{\tilde{\tau}}^2+
  \vert {\rm det}\, \hat g \vert \,   \frac{d\si^2}{\si}\qquad {\rm with} \qquad 
  {\rm det}\, \hat g =  {\lambda^2   \over  {A  (\sigma_- - \sigma)}}\  . 
\end{equation}
The  worldvolume is   timelike for  all $\sigma >  \sigma_-$, as already advertised. 
More interestingly, the metric   (felt by signals that propagate on the brane) 
is that of a two-dimensional black-hole  with  horizon at the ergoplane
 $\sigma=0$. This lies outside the bulk horizons $\sigma_+^{{\rm H}j}$, in agreement  with  
   arguments showing   that the causal structure is always set
 by the Einstein metric \cite{Gibbons:2000xe}. Similar remarks in a closely-related context 
were made before  in ref.\cite{Frolov:1995qp}. 
  The brane-horizon (bH) temperature,  
 \bea
4\pi  \Theta_{\rm bH} =    \big(-{\rm det}\, \hat g\vert_{\sigma=0}  \big)^{-1/2} \ ,  
 \eea
is intermediate between   $\Theta_1$ and $\Theta_2$  as can be easily checked. For $\ell_1=\ell_2$   for example one finds
$ 2\,\Theta_{\rm bH}^2 = \Theta_1^2+ \Theta_2^2$.



 \section{The non-Killing  horizon}\label{sec:6n}

 Since   $\sigma_-$ lies behind an inner  horizon,   the first
singularities of  the embedding functions \eqref{xj} are at    $\sigma_+^{{\rm H}j} $.   
A key  feature of  the non-static solutions is  that  these outer  BTZ  horizons, which are apparent horizons as  will  become clear,
  do not meet  at the same point on the brane. 
 For  $  J_j\not=  0$ the following strict inequalities indeed
hold 
   \begin{equation}\label{orderH} 
   \sigma_+^{{\rm H}1} >  \sigma_+^{{\rm H}2}    \quad {\rm if} \ \ M_1 > M_2\, ; 
\qquad 
  \,  \sigma_+^{{\rm H}2} <  \sigma_+^{{\rm H}1}    \quad {\rm if} \ \  M_1< M_2 \  .  
 \end{equation}  
For small $J_j$ these inequalities are manifest   by Taylor expanding   \eqref{sH},
\bea
 \sigma_+^{{\rm H}j}  = - {J_j^2 \over M_j} + O(J_j^{\,4})\ . 
\eea
 We show  that they hold  for all $J_j$ 
  in appendix \ref{app:B}.

     The  meaning of these inequalities  becomes clear if we use   the holographic dictionary \eqref{J}, the energy currents
     \eqref{4flows} and the detailed-balance condition \eqref{detbal} to write  the $M_j$ as follows
  \begin{equation}\label{theMj} 
    \begin{split}
 &
  M_1 = 2\pi^2 \left[ \Theta_1^2 (1+{\cal R}_1) + \Theta_2^2 (1- {\cal R}_1) \right] \ ;  
\\[1ex]
&
\, \, 
  M_2 = 2\pi^2 \left[ \Theta_1^2 (1-{\cal R}_2) + \Theta_2^2 (1+{\cal R}_2) \right] \ . 
   \end{split}
\end{equation}    
Assuming  $0\leq {\cal R}_j \leq 1$, we see that the hotter side of the interface has the larger $M_j$. 
What  \eqref{orderH}  therefore says is  that the brane  hits the  BTZ  horizon of the hotter side  first.


\subsection{The arrow of time}\label{sec:arrow}

\begin{figure}[thb!]
           \vskip -2cm 
  \centering  
\includegraphics[width=16cm]{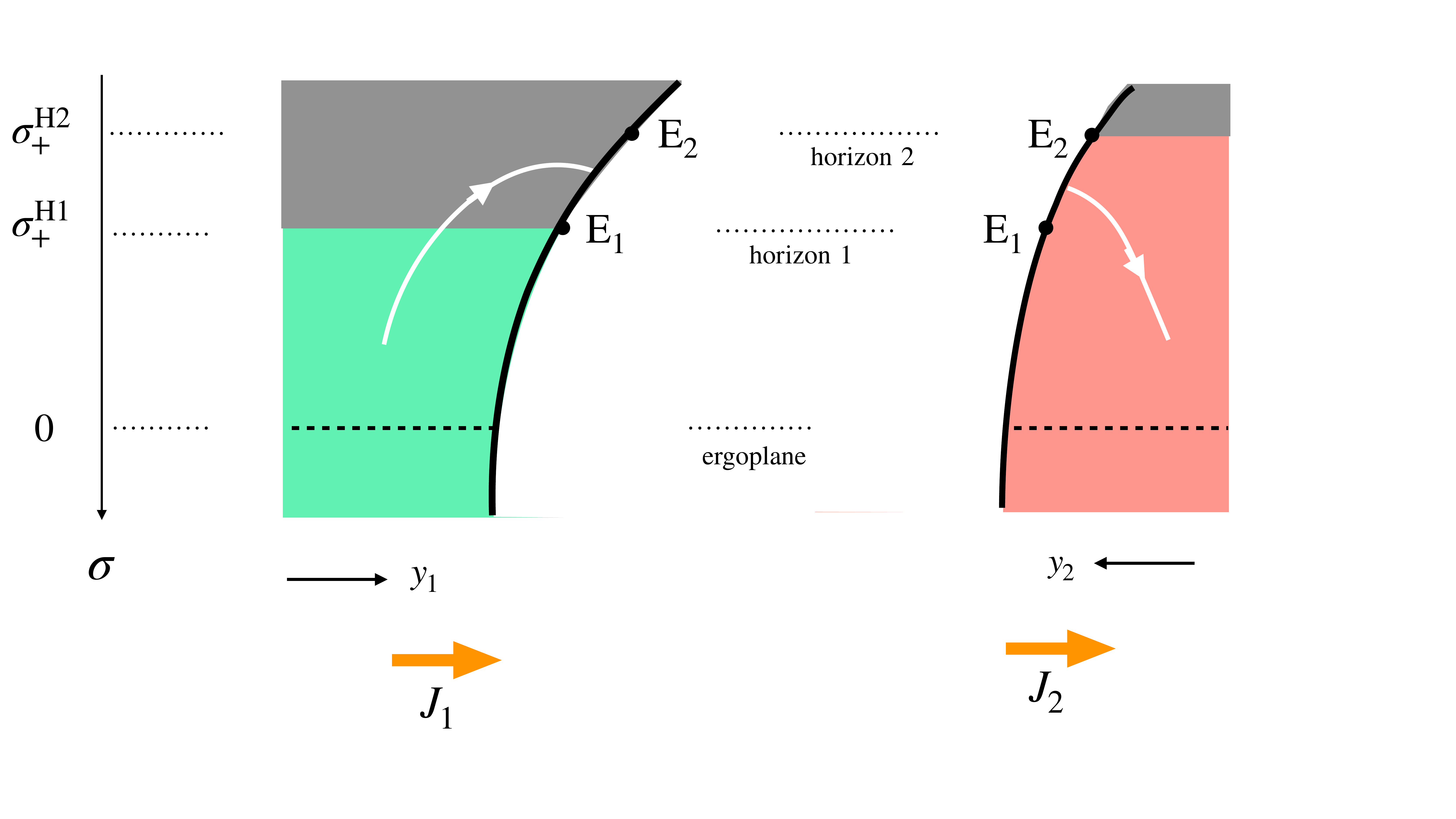}
\vskip -8mm
     \caption{\footnotesize A brane (thick black curve)
     entering  the local outer  horizons  ${\cal H}1$ and ${\cal H}2$ (the boundaries of the grey regions in the figure)
     at two different points  E1 and E2.  
  The  piece [E1,E2] of the wall
     is behind the horizon of slice  1 but outside the horizon of  slice  2. The thick orange arrows show the direction of heat flow.  
  The  white curve is the worldline of an observer entering ${\cal H}1$, crossing the brane and emerging outside 
     ${\cal H}2$. }
\label{fig:5n}
 \end{figure}

Assume for  concreteness  $M_1> M_2$, the case $M_2>M_1$ being  similar.\footnote{Strictly speaking we  
also ask  that the  brane hits both outer horizons   before the  inner (Cauchy) horizons, since we cannot trust our classical solutions
beyond the  latter. As explained in  appendix   \ref{app:B}, this condition is automatic when $M_2>M_1$, but not 
when $M_1>M_2$, where it is possible for  some range of parameters  to have $\sigma_+^{{\rm H}2} <  \sigma_-^{{\rm H}1} $.
}
From eq.\eqref{s+0} we have  $M_1 = M_2 + \lambda \vert J_1\vert$. We do not commit yet  on the sign of $J_1$, nor on 
the   sign in eq.\,\eqref{s+0}, but the product of the two should be  positive. 
 Figure  \ref{fig:5n} shows the behaviour  
of the brane past the ergoplane. 
The vertical axis
  is  parameterised by  $\sigma$
 (increasing  downwards), and  the horizontal axes by  the 
 ingoing Eddington-Finkelstein  coordinates $y_j$ defined
 in eq.\,\eqref{EFc}. These coordinates  are regular at the future   horizons, and  reduce to the flat  ICFT coordinates $x_j$ at the 
AdS   boundary. 
 
 \smallskip

 Let us take a closer look  at the  wall embedding in Eddington-Finkelstein (EF) coordinates. 
From   eqs.\,\eqref{xj}   and  the identities 
 $r_j' = 1/2r_j$\,  we get  
  \begin{equation}\label{yj} 
    \begin{split}
 &
 y_1' =    \frac{\ell_1} {2(\sigma- \sigma_+^{\rm H1} )(\sigma- \sigma_-^{\rm H1})}
 \left[ 
     {J_1\ell_1 \over 2 \sqrt{\sigma + M_1\ell_1^2}}
  - \,  \frac{(\lambda^2+\lambda_0^2)\,\sigma  \, + \lambda \vert J_1\vert } { 
    \sqrt{A (\sigma - \sigma_-) }} 
 \right] \ , 
\\[1ex]
&
\, \,  y_2' =     \frac{\ell_2} {2(\sigma- \sigma_+^{\rm H2} )(\sigma- \sigma_-^{\rm H2})}
 \left[ 
     {J_2\ell_2 \over 2 \sqrt{\sigma + M_2\ell_2^2}}
  - \,  \frac{(\lambda^2- \lambda_0^2)\,\sigma  \, -  \lambda \vert J_2\vert } { 
    \sqrt{A (\sigma - \sigma_-) }} 
 \right] \ . 
    \end{split}
\end{equation}    
A little algebra shows that  the square brackets in the above expression  vanish  at the corresponding horizons
$\sigma= \sigma_+^{{\rm H}j}$ if   $J_1=-J_2>0$. 
 The  functions $y_j$ are in this case  analytic at the horizons. 
By contrast, if  $J_1=-J_2 <0$ these functions are  singular:  $y_1 \to +\infty$ at $\sigma_+^{{\rm H}1}$,\, and
$y_2 \to - \infty$ at $\sigma_+^{{\rm H}2}$. We interpret this 
 as evidence that $J_1$ must  be positive, 
as expected from  the arrow of  heat flow in the boundary ICFT. This means that $M_1=M_2+\lambda J_1$, and hence 
  the sign in  the  expression  \eqref{Tj1} for the 
transmission coefficients is also plus, 
 in agreement with the result of ref. \cite{Bachas:2020yxv}.

 Note that time reversal flips  the sign of  the $J_j$ and   leaves $M_j$ unchanged. Since time reversal is a symmetry of the
 equations, both 
  signs of $J_1$ give therefore solutions --  one diverging  in the past and the other  in the future horizons.  
One can check for consistency that choosing the minus sign in the expressions \eqref{Tj1} interchanges the
incoming  and outgoing energy currents in \eqref{4flows}.
  Similarly to  a white hole,   which solves  Einstein's equations but cannot
be produced by gravitational collapse, we  expect  that  no physical protocol  can prepare  the   $J_1<0$ solution.
\smallskip


\subsection{Event versus apparent horizon}\label{sec:62n}

   Denote by  ${\cal H}_1$  and ${\cal H}_2$  the horizons  of the two BTZ regions of the 
   stationary geometry, and by  {\small E}$_1$ and  {\small E}$_2$
  their  intersections  with the brane worldvolume. We can  foliate spacetime by Cauchy slices  $v_j = \bar v + \epsilon_j(r_j,x_j)$,
    where $\bar v$ is a uniform foliation parameter.\footnote{The non-trivial radial dependence in the definition of the 
    Cauchy slice  is necessary  because 
constant  $v_j $ curves are   lightlike behind the $j$th horizon.}
    We use the same symbols for the  projections of  ${\cal H}_j$ and {\small E}$_j$
      on a   Cauchy slice. Since simultaneous translations of $v_j$ are Killing
isometries,  the  projections do not depend on $\bar v$.

 Both  ${\cal H}_1$  and ${\cal H}_2$ are local (or apparent)  horizons, i.e. future-directed  light rays can only traverse them in one direction. 
 But it  is clear  from  figure \ref{fig:5n}   that ${\cal H}_1$ cannot be part of the  event horizon of global spacetime. 
 Indeed,  after  entering   ${\cal H}_1$ an observer moving to the right can 
 traverse  the  [{\small E}$_1$, {\small E}$_2$] part of the wall,   emerge outside ${\cal H}_2$
in region  2, and from there   continue her journey to the boundary.  
Such  journeys are  only forbidden if  {\small E}$_1$= {\small E}$_2$,
i.e. for the  static equilibrium solutions.

    In order to analyse the problem  systematically, we  define
     an  everywhere-timelike unit vector field  that distinguishes the past from future,  
     \bea
     t^\mu \partial_\mu  = {\partial \over  \partial v_j} + \, {h_j(r_j) -1\over 2\ell_j}\, {\partial \over  \partial r_j}+ 
      {J_j\ell_j\over 2r_j^2} \,  {\partial \over  \partial y_j}
      \qquad {\rm in \ the \  {\it j}th\ region\,. } 
    \eea 
 Using the metric \eqref{EFmetric} the reader can check that $t^\mu t_\mu = -1$. 
 To avoid charging the  formulae we  drop temporarily  the  index $j$. 
  A future-directed null curve  has tangent  vector
\bea\label{nullrays}
\dot x^\mu = ( \dot v , \dot r , \dot y )\   \qquad {\rm where} \quad \dot x^\mu\dot x_\mu = 0 \ \ \ {\rm and} \ \ \ 
\dot x^\mu t_\mu   < 0 \ . 
\eea 
The  dots denote  derivatives  with respect to a  parameter  on  the curve.
Solving the conditions \eqref{nullrays} gives 
\bea\label{nullray}
     \dot r = {h\over 2\ell}  \dot v   - {r^2\over 2\ell \dot v}  \bigl(\dot y -  {J\ell\over 2r^2} \dot v \bigr)^2\  \quad  {\rm and} \ \  \ 
\dot v   > 0 \ . 
\eea 
 We see that the arrow of time is defined by  increasing $v$,  
   and that behind the horizon, where $h(r)$ is negative,    $r$ is   monotone decreasing
 with time. 
   This suffices to show that ${\cal H}_2$ {\it is}  part of the event horizon  -- an observer 
  crossing it  will  never make it out to  the boundary  again.

\begin{figure}[thb!]
 \centering  
\includegraphics[width=16cm]{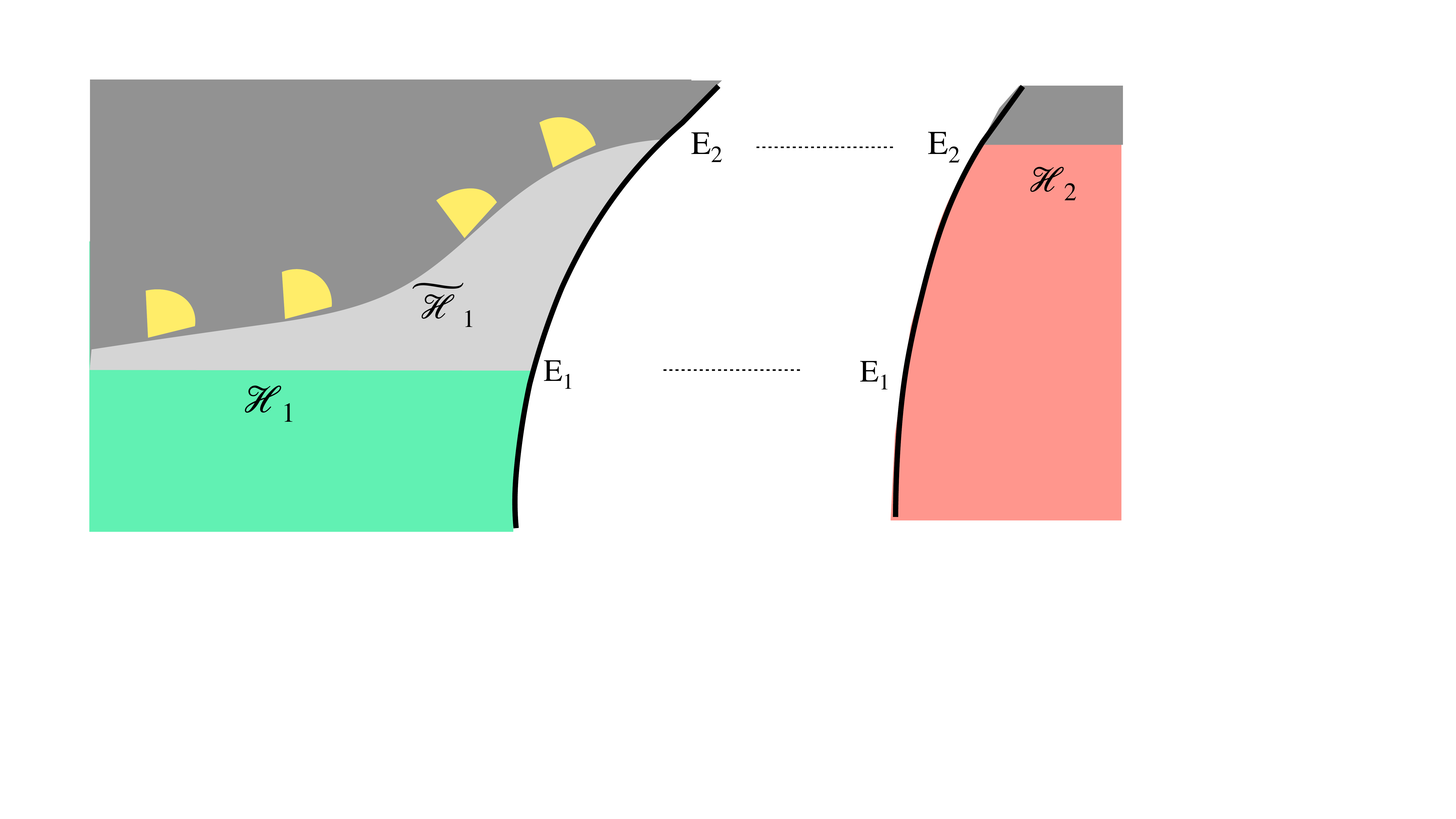}
\vskip -25mm
     \caption{\footnotesize The event and apparent horizons,
     $\widetilde {\cal H}_1  \cup {\cal H}_2$ and $ {\cal H}_1  \cup {\cal H}_2$, as 
        described in the text. 
     The event horizon   is connected but it is not Killing.   Projections of the  local  light-cone  
     on a Cauchy slice  are shown in yellow.
     The  light grey region behind  ${\cal H}_1$ is outside the event horizon because
   signals can escape towards  the right.
    }
\label{fig:6n}
 \end{figure}

   As explained above, the story differs in region 1. Here  the event horizon consists of a 
  lightlike surface $\widetilde{\cal H}_1$  such that  no future-directed causal curve starting from a point behind it  can  reach the 
  [{\small E}$_1$,{\small E}$_2$] part of the wall. Clearly, the  global event horizon 
 \bea\label{eventH}
  {\cal H}_{\rm event} =  \widetilde{\cal H}_1\, \cup\, {\cal H}_2
  \eea
 must be  continuous   and lie  behind the apparent horizon ${\cal H}_1$ in region 1. This is  illustrated in figure \ref{fig:6n}. 
   General theorems \cite{Hawking:1973uf} actually show   that a local  horizon  which is  part of
  a  trapped compact surface cannot lie outside  the event horizon.  But there is no clash with these theorems here
  because   ${\cal H}_1$
  fails to be  compact, both  at infinity and  at  {\small E}$_1$.

To compute the projection of  $\widetilde{\cal H}_1$ on a  Cauchy slice,  note that  
  it is a curve through  the point  {\small E}$_2$ that is   everywhere tangent to
  the projection of the  local light cone,  as  shown  in the figure. Put differently, at every point on the curve we 
  must   minimise  the angle between (the projection of)   light-like vectors  
    and  the positive-$y_1$ axis. 
    This will guarantee  that  an   observer starting behind $\widetilde{\cal H}_1$ will not  be able to 
    move fast enough towards   the right in order  to hit 
    the wall before the point  {\small E}$_2$.

    Parametrising  the curve by $y_1$,  using   
     eq.\,\eqref{nullray}  and dropping again for simplicity the $j=1$ index we find 
        \bea\label{nullra}
   - {dy\over dr}\,\bigg\vert_{\widetilde {\cal H}_1} \ = \   
  {\rm max}_{\, v_y >0}\ 
 \bigg[  {r^2\over 2\ell v_y}  \bigl(1  -  {J\ell\over 2r^2} {v_y }  \bigr)^2 - {h\over 2\ell}  {v_y}   \bigg]^{-1}   \ , 
     \eea 
  where $v_y  \equiv dv/dy$. The extrema of this expression  are $v_y   = \pm r/\sqrt{M\ell^2 - r^2}$. 
  Recall  that we are interested in the region behind the BTZ  horizon  and
  in  future-directed light rays for which   $v$ is monotone increasing  (whereas  
      $r$ is monotone decreasing). 
     For  null rays moving to the right we should thus pick  the positive $v_y$ extremum. 
     Inserting   in \eqref{nullra}   gives the differential equation obeyed by  $\widetilde {\cal H}_1$, 
       \bea\label{nullr}
   {dy\over dr }\,\bigg\vert_{\widetilde {\cal H}1 } \ = \  
   { 2 \ell \over J\ell -  2 r  \sqrt{  M\ell^2 -r^2 } } 
\     \  .  
     \eea 
The (projected) event horizon in region $j=1$ is the integral of 
  \eqref{nullr} with  the  constant of integration fixed so that  the curve passes through   {\small E}$_2$. 
   
  Here now comes the important point. The reader can check that 
near the BTZ  horizon, $r =  r_+^{\rm H1}(1+ \epsilon)$ with $\epsilon\ll 1$, the denominator in \eqref{nullr} 
vanishes like $1/\epsilon$.  This is a non-integrable singularity, so 
  $y(r)$ diverges at $r_+^{\rm H1}$ and hence    ${\widetilde {\cal H}1}$
  approaches   asymptotically   ${\cal H}1$ as  announced in section \ref{sec:3.2n}. The 
  holographic coarse-grained entropy 
   will therefore asymptote to that of the  equilibrium  BTZ horizon, given by 
  eqs.\,\eqref{qen} and \eqref{qen1}. 
 This shows   that the chiral outgoing fluid is  
  thermal, not only in the cold region 2 but also in the hotter region 1.

  

\subsection{Remark on  flowing  funnels}
\label{sec:61n}

\begin{figure}[thb!]
   \vskip  -7mm
 \centering  
\includegraphics[width=15cm]{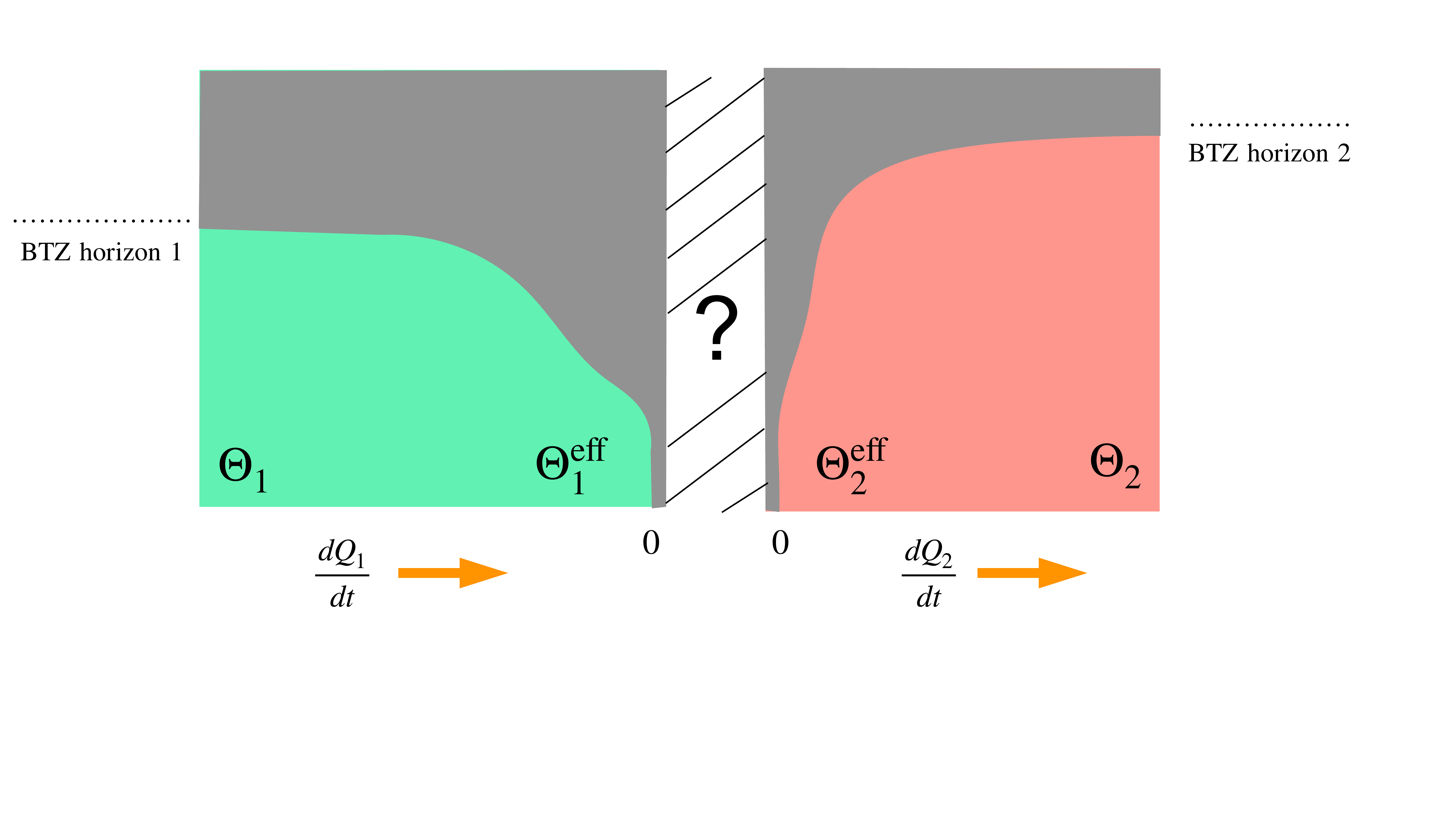}
\vskip -18mm
     \caption{\footnotesize A two-sided flowing   funnel that can mimic  the  energy and entropy flows  of  the holographic interface. 
Tuning the horizon temperatures  so that  the  boundary black hole does not absorb any energy is,  
however, an adhoc condition.  }
\label{fig:bfun}
 \end{figure}

       The fact that  outgoing fluxes are thermalised means that, in what concerns the  entropy and energy flows, 
        the interface behaves like  a  black cavity. This latter can be modelled by a non-dynamical, two-sided  boundary black hole 
        whose (disconnected)  horizon  consists of two points. 
           To mimic the behaviour of the  interface, the two horizon temperatures should be equal to the  $\Theta_j^{\rm eff}$ that
         saturate the   bounds \eqref{boundsS}. This is illustrated 
         in  figure \ref{fig:bfun}.                     

The precise shape  of the flowing horizon(s)  depends on the  
 boundary  black hole(s)  and is not important for our purposes here.  For completeness,  following
 ref.\cite{Fischetti:2012ps},  we outline  how to derive it 
  in appendix \ref{app:C}.  Like   the thin-brane horizon  of figure \ref{fig:6n},  it approaches the BTZ horizons at infinity 
  but differs   in the central region (notably with a delta-function peak in the entropy density at $x= 0$, see appendix \ref{app:C}). 
  
   The key difference is however elsewhere. 
The two halves of the flowing funnel of  figure \ref{fig:bfun} are a priori  separate solutions, with the  
 temperatures $\Theta_j$ and $\Theta_j^{\rm eff}$ 
  chosen at will.   But to mimic the conformal interface one  must impose  continuity of the heat flow, 
\bea\label{heatfun}
{dQ_1\over dt}  = {\pi c_1 \over 12} (\Theta_1^2 -  (\Theta_1^{\rm eff})^2)  =  {\pi c_2 \over 12} ( (\Theta_2^{\rm eff})^2 - \Theta_2^2) =  
{dQ_2\over dt} \ . 
\eea
This relates the  horizon  temperatures   to each other and to those of the distant heat baths. 
  It is however unclear  whether any  local condition behind the event horizons can impose the  condition
  \eqref{heatfun}\,.

  
    \section{Pair of interfaces}\label{sec:7n}
  
  In this last section we  consider  a pair of identical 
   interfaces between  two theories, CFT$_1$ and CFT$_2$.\footnote{  
  Our branes are not oriented, so there is no difference between an interface and  anti-interface. More general
  setups  could include several different  CFTs and triple   junctions of branes, but such systems  are beyond the scope of the present work.
   }  
   The interface  separation is  $\Delta x$.    Let the theory that lives  in the finite interval be  CFT$_2$
  and the theory outside  be CFT$_1$ (recall that we are assuming $\ell_2 \geq \ell_1$). 
    At   thermal equilibrium the system undergoes a first-order  phase  transition  at a critical
    temperature $\Theta_{\rm cr} = b/\Delta x $ where $b$ depends on 
    the classical Lagrangian parameters $\lambda \ell_j$   \cite{Bachas:2021fqo, Simidzija:2020ukv}. 
 Below $\Theta_{\rm cr}$ the brane avoids  the horizon and is  connected, while 
     above $\Theta_{\rm cr}$ it breaks into two disjoint pieces   that hit separately the singularity of the black hole. 
    This   is  a variant of the  Hawking-Page phase transition  \cite{Hawking:1982dh} that  
        can  be interpreted  \cite{Witten:1998zw}
    as a deconfinement transition of  CFT$_2$.

\begin{figure}[thb!]
   \vskip  -2mm
 \centering  
\includegraphics[width=16cm]{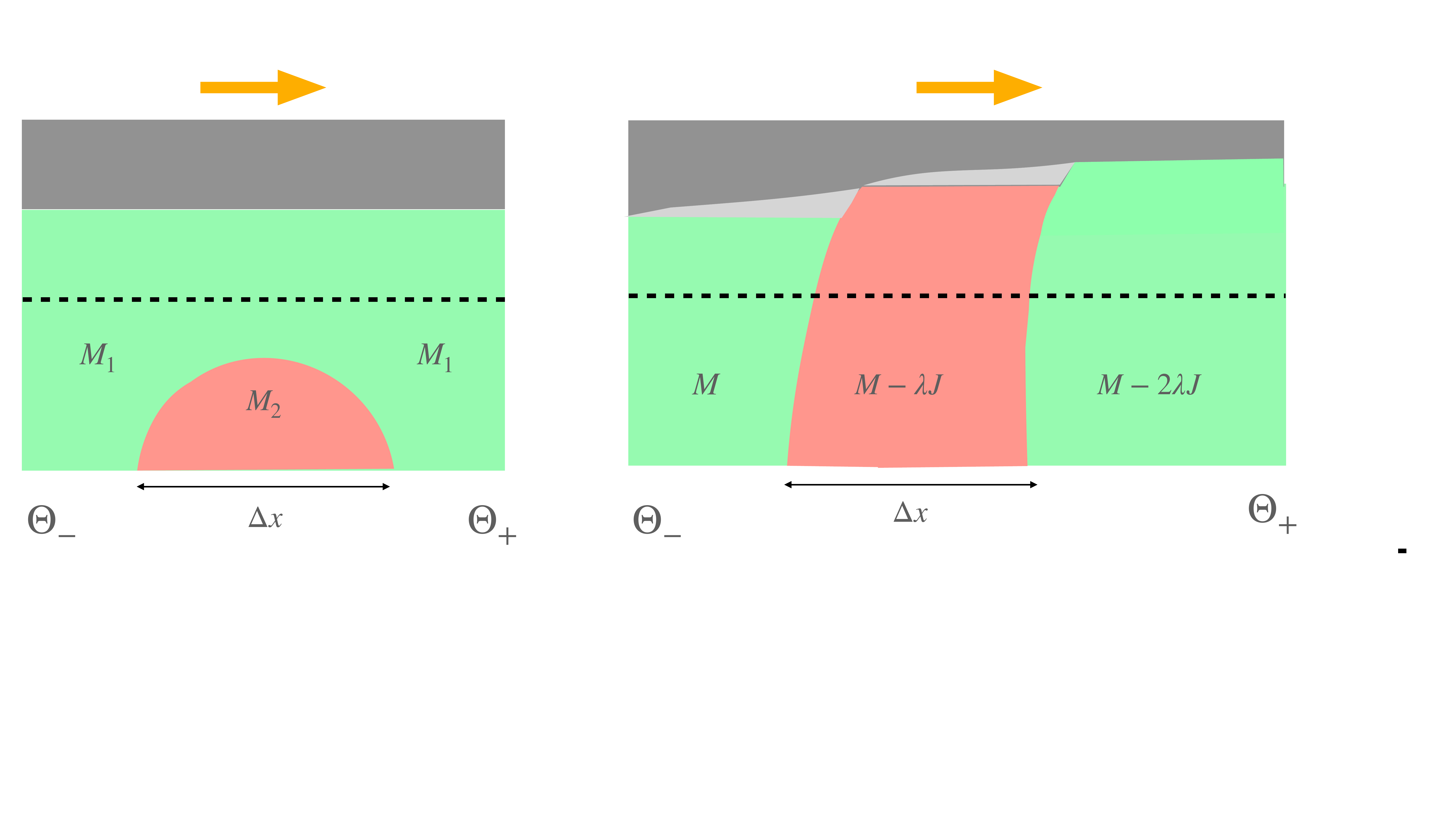}
\vskip -26mm
     \caption{\footnotesize The  two types of NESS  for an interface pair. In the 'quantum phase' (left)
     heat conducts as if there was no scatterer, while in the 'classical phase' (right)   the  conductance is the same 
      as for  an isolated CFT$_1$ defect dual to 
      a  brane of  tension $2\lambda$. Also given  in the figure is the  BTZ mass parameter in  different  regions of the geometry.
      The yellow arrows show the direction
      of heat flow. 
    }
\label{fig:7n}
 \end{figure}

  {   We would like to understand what happens when this  system is  coupled  to  reservoirs  
     with  slightly different  temperatures $\Theta_\pm =\Theta \pm d\Theta$
     at $x=\pm \infty$.  Because of the temperature gradient   the
      branes are now stationary, but  they conserve the topology  of their static ancestors. In the low-$\Theta$ phase 
    the   brane    avoids  the  
    ergoregion (which is displaced from the horizon infinitesimally) and stays connected,  while in the high-$\Theta$ phase it splits in 
    two disjoint   branes that  enter the ergoregion and  
 hit  separately  a Cauchy horizon or a bulk singularity. 
 The   two phases  are   illustrated  
   in figure \ref{fig:7n}. 
   }

   Consider the  high-$\Theta$ phase first. The    isolated-brane solution  of  sections 
   \ref{sec:5n} and \ref{sec:6n}  is  here juxtaposed to    a solution  in which the roles  of
   CFT$_1$ and CFT$_2$ are inverted.  The mass parameter  of the three BTZ regions 
   decreases  in the direction of heat  flow, jumping by $ \lambda  J$ across each brane.
   This is indeed the `ticket of entry'  to the ergoregion, as explained in  eq.\,\eqref{s+0}  and  section \ref{sec:arrow}.  
     The total change of BTZ mass  across the pair  is the same as if the two branes had merged into a single one
    with twice  the tension.  Using  the holographic dictionary \eqref{J}  and the fact that incoming fluxes at $x=\pm\infty$ are thermal 
   with  temperatures  $\Theta_\pm$  one indeed  computes
   \bea\label{prob1}
  \underline{ {\rm high}\ \Theta} :  \qquad  {dQ\over dt}    \,=\, {\pi^2 \ell_1 \over 1+ \lambda \ell_1} ( \Theta^2_- - \Theta^2_+)\ 
    \equiv \  {\pi^2 \ell_1  }{\cal T}_{\rm pair}  ( \Theta^2_- - \Theta^2_+)\, . \ \ \ 
   \eea
where the effective  transmission coefficient ${\cal T}_{\rm pair} $ is  that of  a  CFT$_1$ defect whose 
dual  brane has  tension  $2\lambda$. Note in passing 
that this  effective  brane tension   can exceed the upper bound
\eqref{3crtens} above which an individual   brane inflates, and that an array of widely-spaced  branes can make the transmission coefficient arbitrarily small. 
   
   The heat flow   \eqref{prob1} is what one would obtain from   classical scatterers.\footnote{The argument grew out  of
           a conversation  with   Giuseppe Policastro who noticed that the  tensions of two juxtaposed branes 
           effectively add up in  the calculation of ref.\cite{Bachas:2020yxv}. } 
         To understand    why, think of   ${\cal T}_j$ 
         and  ${\cal R}_j$ as   classical
         transmission and reflection probabilities  for quasi-particles incident on the  interface 
         from the side $j$. The probability of passing  through both interfaces
        is the sum of probabilities  of trajectories with any number   of double reflections in between, 
         \bea\label{prob} 
          {\cal T}_{\rm pair}=  {\cal T}_1\, ( 1 +    {\cal R}_2^2 +  {\cal R}_2^4  + \cdots  ) \, {\cal T}_2 \,  
           = \, { {\cal T}_1 {\cal T}_2  \over 1 -   {\cal R}_2^2 } =  {1\over 1+ \ell_1 \lambda}  \ , \ \ \ 
          \eea
         where in the last step we used   the holographic relations \eqref{Tj1}.  This gives precisely  
         the result \eqref{prob1} as advertised. 
          
             The low-$\Theta$ case  is drastically different.  The solution   is now obtained   by  
             gluing   a brane
             with  a turning point (i.e. $\sigma_+>0$, see  section \ref{sec:4.2n})  to
                its mirror image,  so that the brane has reflection symmetry. The bulk metric,  however, is not $\mathbb{Z}_2$
                symmetric because in the mirror image we do not flip the  sign of   the BTZ `spin'   $J$. 
                 This is required for  continuity of the $dx dt$ component of the bulk metric, and it is 
                   allowed because  when the brane avoids  the ergoregion there is no regularity condition to fix 
             the sign of $J$, as in section \ref{sec:62n}. 
                 The  BTZ mass 
               is  thus  the same at $x= \pm\infty$,  while its value in the CFT$_2$ region depends on the interface
               separation $\Delta x$. It 
                follows   from the holographic dictionary \eqref{J} that 
                the heat flow is  in this case  unobstructed,  
                 \bea\label{quan}
                  \underline{ {\rm low}\ \Theta} : \qquad 
    {dQ\over dt}    \,=\,   {\pi^2 \ell_1  } ( \Theta^2_- - \Theta^2_+)\, ,  
   \eea
                i.e. the effective transmission coefficient is ${\cal T}_{\rm pair} =1$.  
                  Superficially,  it looks  as if  two branes with equal and opposite tensions  have merged into a  tensionless one.

                 In reality, however, the above phenomenon   is 
                  deeply quantum. What the   calculation says is that when a  characteristic 
             thermal wavelength  becomes larger than    the interface 
               separation,   
               coherent scattering  results in   all incident energy being   transmitted. 
                This is all the more  surprising since   CFT$_2$ is in the confined  phase, and one could  have expected 
                that   fewer degrees of freedom  are available to conduct heat. 
                  The microscopic mechanism behind
                this surprising   phenomenon deserves to be  studied further.

       The above discussion stays valid for finite temperature difference $\Theta_+-\Theta_-$,   but the dominant phase
       cannot in this case be found  by comparing free energies. Nevertheless, as $\Delta x\to 0$ we expect from the dual ICFT
       that the interface-antiinterface pair fuses into the trivial (identity) defect \cite{Bachas:2007td}, whereas at very large $\Delta x$ the connected solution ceases
       to exist. A transition is therefore bound to occur   between these  extreme separations.

            Let us comment finally on what happens if the interval theory is CFT$_1$, the theory with fewer degrees of freedom,
            and the outside theory is CFT$_2$.  Here the low-temperature
            phase only exists for sufficiently-large   tension if  $c_1< c_2<3 c_1$, and 
             does not exist  if   $c_2> 3 c_1$ 
             \cite{Bachas:2021fqo, Simidzija:2020ukv}.  
           The (sparse) degrees of freedom of
       the  interval  theory in this latter case are always in the high-temperature phase,
        and there can be no quantum-coherent conduction of heat.
           Reassuringly, this  includes the  limit $c_2/c_1 \to 0$ in which  the CFT$_1$ interval is effectively void. 
           
        Note also that in the low-temperature  phase the wire can be compactified to a circle and the heat current can be sustained
        without external reservoirs. This is  not possible in the high-temperature phase.


   \section{Closing  remarks}\label{sec:8n}  
  
    The study of far-from-equilibrium  quantum systems is   an exciting frontier both 
    in condensed-matter
    physics and 
    in quantum gravity. Holography is a   bridge between these two areas of research, and  has 
    led to  many new insights. Much remains however to be understood, and simple 
    tractable models can help as 
    testing grounds for new ideas.  
     The holographic NESS  of this paper are   tractable thanks to several simplifying factors:
   2d  conformal symmetry,  isolated impurities and the assumption of a thin brane.  If the first two can be justified in  (very)
   pure ballistic systems, 
    the thin-brane approximation is an adhoc assumption of convenience.  
        Extending our  results    to  top-down  dual pairs  
  is  one urgent open question.

      Another obvious question concerns  the structure of entanglement and the  Hubeny-Rangamani-Ryu-Takayanagi  curves
    \cite{Ryu:2006bv,Hubeny:2007xt} 
       in the above  steady states. While  it is known that geodesics  cannot probe  the region behind  equilibrium  horizons 
       \cite{Hubeny:2012ry}, they can reach behind both apparent and event horizons in  time-dependent backgrounds, 
        see e.g.  \cite{AbajoArrastia:2010yt,Balasubramanian:2010ce,Hartman:2013qma}.   
    In the framework of the  fluid/gravity correspondence the   entropy current  associated to the event horizon is a 
    local functional of the boundary data \cite{Bhattacharyya:2008xc}. 
    It would be interesting to examine this question  in the present far-from-equilibrium context. 
       Note also that the particularly simple form of   matter in our problem (a thin fluctuating brane) may allow
       analytic calculations of the quantum-corrected extremal surfaces
  \cite{Engelhardt:2014gca,Faulkner:2013ana}.

    Another interesting question  is  how  the deconfinement transition of the interval  CFT 
    in  section \ref{sec:7n} relates   to  the sudden jump in   thermal conductivity of the system. 
    Last but not least,  it would be nice to relate the production of coarse-grained entropy to the scattering matrix of 
     microscopic interfaces, e.g.  for  the simplest free-field interfaces of 
    \cite{Bachas:2001vj,Bachas:2007td,Bachas:2012bj}. 
   
    We hope to return to some of these questions  in the near future.


 \vskip 0.4cm

 {\bf Aknowledgements}: 
 We  thank  Denis Bernard, Shira Chapman, Dongsheng Ge, Andreas Karch,   Joao Penedones  and Giuseppe Policastro
 for  many stimulating discussions
 during the course of this work. 
 We are indebted  to Giuseppe Policastro for the  observation  that led to   
  the argument  around equation \eqref{prob}.  C.B. also aknowledges 
the support of the NYU-PSL  project ”Holography and Quantum Gravity” (ANR-10-IDEX-0001-02 PSL).

  \vskip 0.4cm

 {\bf Note added}:  
In more than 1+1  dimensions the two shock waves of  figure \ref{fig:protocols} 
are replaced by a shockwave on the cold side   and a broadening  rarefaction wave on the hot side, see  \cite{Lucas:2015hnv,Spillane:2015daa,Ecker:2021ukv}.  Although this  only affects  the approach 
to  the NESS in the partitioning protocol,  not the NESS proper, it is interesting  that  
 the event horizon of our interface NESS follows a similar pattern: it coincides with the  equilibrium BTZ  horizon on the colder side
  but approaches it only asymptotically on the hotter side. 
We thank
  Julian Sonner,  and  also  Christian Ecker,  Johanna Erdmenger and Wilke Van Der Schee for email exchanges on this point.
 

 

 


\appendix

\section{Solving the thin-brane equations}\label{app:A} 

From the form \eqref{metric} of the bulk metric and the  embedding ansatz \eqref{ansatz} of  a stationary brane,  we derive  the 
following  continuity equations   for the induced  metric
\begin{equation}\label{A1} 
  \hat g_{\tau\tau} \, =\,   M_1\ell_1^2 -  r_1^2 = M_2\ell_2^2 - r_2^2\,,
\end{equation}
\begin{equation}\label{A2} 
   \hat g_{\tau\sigma} \, =\,     (M_1\ell_1^2- r_1^2)f_1' -  {J_1\ell_1\over 2}  x_1' =  (M_2\ell_2^2 - r_2^2)f_2'  -  {J_2\ell_2\over 2}  x_2'\,,
\end{equation}
\begin{equation}\label{A3} 
    \begin{aligned}{}
      \hat g_{\sigma\sigma} \, =\,  &\frac{\ell_1^2\,r_1'^2}{h_1(r_1)}+r_1^2x_1'^2-J_1\ell_1 x_1'f_1'
      + (M_1\ell_1^2- r_1^2)f_1'^2\\
        &=\, \frac{\ell_2^2\,r_2'^2}{h_2(r_2)}+r_2^2x_2'^2-J_2\ell_2 x_2'f_2'+(M_2\ell_2^2 - r_2^2)f_2'^2\,.
    \end{aligned}
\end{equation}
 The primes denote  derivatives  with respect to $\sigma$, and  the function $h(r)$ has been  defined in 
 eq.\eqref{metric2}, 
 \bea
 h(r)\, =\,  {r^2 } - M\ell^2  + {J^2 \ell^2\over 4r^2} \, =\,  {1\over r^2} (r^2 - r_+^2)(r^2-r_-^2)  \ . 
 \eea

Following ref.\cite{Bachas:2021fqo}  we
choose  the convenient parametrization $ \sigma = - \hat g_{\tau\tau}$\,,  so that   $r_j^2 = \sigma + M_j\ell_j^2$\, and 
$r_j' = 1/2r_j$. This parametrization
need not be one-to-one,  
it  actually only covers    half  of the   wall when  this latter  has a turning point.  With this choice 
the ergoplane is located at 
$r_j^2 = M_j \ell_j^2\, \Longrightarrow\, \sigma =0$, and the functions $h_j$ can be written as 
\begin{equation}
    h_j(\sigma) =    {\sigma^2 + \sigma  M_j\ell_j^2 + J_j^2\ell_j^2/4 \over \sigma + M_j\ell_j^2}\ =\
   { (\sigma -  \sigma_+^{{\rm H}j} )(\sigma - \sigma_-^{{\rm H}j} )  \over \sigma + M_j\ell_j^2}  \ , 
\end{equation}
where
\bea\label{sHpm}
\sigma_\pm^{{\rm H}j} =  - {M_j\ell_j^2\over 2}   \pm  {1 \over 2}\sqrt{M_j^2 \ell_j^4 -J_j^2 \ell_j^2 } \  
\eea  
are  the  locations of the  horizons in the $j$th chart.
\smallskip

  From \eqref{A1}\,-\,\eqref{A3} one computes  the determinant of the induced metric 
  \begin{equation}\label{det} 
   -  {\rm det}(\hat g)  
   \,  = \,   \frac{\sigma \ell_j^2}{4r_j^2 h_j} + {h_j r_j^2} x_j'^{\,2}   \,.  
\end{equation}
Note that it  does not depend on  the time-delay functions $f_j(\sigma)$, because 
 these   can be absorbed by  the  unit-Jacobian
reparametrization  
$$\tilde\tau = \tau + f_j(\sigma), \quad \tilde \sigma = \sigma\,.$$ 
Eq.\,\eqref{det} can be used to express  the $x_j^\prime$ (up to a sign)   in terms of  det\,$\hat g$. 
 A combination of  eqs.\,\eqref{A1} and \eqref{A2} expresses, in turn,   the time delay across the wall in terms of the
  $x_j'$, 
\bea\label{A7} 
 \sigma (f_2^\prime - f_1^\prime) = {1\over 2} (J_1\ell_1 x_1^\prime -  J_2\ell_2 x_2^\prime)\ . 
\eea
 To complete the calculation we need therefore  to  find  det\,$\hat g$ and then solve the  equations \eqref{det}
 for $x_j'$. 
 

\subsection*{The Israel-Lanczos  conditions}
 
This is done with the help of   the  Israel-Lanczos   conditions \cite{Lanczos,Israel:1966rt}
(see also \cite{Papapetrou})
 which express the discontinuity
of the extrinsic curvature across the wall, eqs.\eqref{IsLan}. We follow the conventions of ref.\cite{Bachas:2021fqo}: 
$K_{\alpha\beta}$ is  the covariant derivative of the inward-pointing unit normal vector,
 and the orientation is such  that for inceasing  $\sigma$   the wall  encircles {\it clockwise} 
the interior of both  charts
 in the $(x_j, r^{-1} )$  planes.\footnote{In these conventions the boundary ICFT  is  folded, with both  
CFT$_1$ and CFT$_2$ living  on the 
 same side of the interface.}

  A somewhat  tedious but straightforward calculation gives
 \begin{equation}
   K_{\tau\tau} =  - \frac{h r^2x'}{ \ell  \sqrt{\vert \hat g \vert } } \qquad {\rm and} \qquad 
      K_{\tau\sigma} =    \frac{h r^2x'}{ \sigma \ell  \sqrt{\vert \hat g\vert }} \,\,\hat g_{\tau\sigma} +  \frac{ J \sqrt{\vert \hat g\vert } }{2\sigma}\ ,  
\end{equation}
where $\hat g$  is  a shorthand notation for  ${\rm det}(\hat g)$. 
      The Israel-Lanczos  equations \eqref{IsLan}  thus read
\begin{equation}\label{Isr1} 
 {1\over \sqrt{\vert \hat g}\vert } \Bigl(  \frac{h_1r_1^2x_1'}{ \ell_1 }+\frac{h_2r_2^2x_2'}{ \ell_2 }\Bigr) 
 = - \lambda  \sigma \,,
 \end{equation}
 \begin{equation}\label{Isr2} 
  {1\over  \sqrt{\vert \hat g}\vert } \Bigl(   \frac{h_1r_1^2  x_1'}{\ell_1 } + \frac{h_2r_2^2  x_2'}{\ell_2  }\Bigr)\,\hat g_{\tau\sigma}
         +  \frac{ \sqrt{\vert \hat g}\vert }{2 }(J_1+J_2)  =  - \lambda \sigma  \,\hat g_{\tau\sigma}
   \,.
\end{equation}   
These  are compatible if and only if 
 \bea\label{cons}
J_1+J_2 = 0 \ ,  
\eea
which translates to energy conservation in the boundary CFT. 
 We have checked  that the third  equation, $[K_{\sigma\sigma}] = - \lambda 
\hat g_{\sigma\sigma}$\,,  is   automatically obeyed and thus redundant.  
As  expected,  by virtue of the momentum constraints the three Israel-Lanczos equations  \eqref{IsLan} reduce to  a single  independent
one  plus  the  ``constant-of-integration'' condition  \eqref{cons}.


\subsection*{The general solution} 

Squaring   twice \eqref{Isr1}  and using \eqref{det} to eliminate the $x_j^{\prime\,2}$\,  leads to  a 
quadratic equation for the determinant. This has a singular solution $ {\rm det} (  \hat  g ) =0$, and a 
   non-pathological  one 
   \begin{equation}\label{B1}
 -{\rm det} (  \hat  g ) = \lambda^2 \sigma^3  \bigg[\frac{4h_1h_2r_1^2r_2^2}{ \ell_1^2\ell_2^2}
    - \big(\frac{h_1r_1^2}{ \ell_1^2} + \frac{h_2r_2^2}{ \ell_2^2} - \lambda^2\sigma^2  \big)^2 \bigg]^{-1}.
\end{equation}
  Inserting  the   expressions for $r_j(\sigma)$ and $h_j(\sigma)$ 
  leads  after some algebra to
   \begin{equation}\label{det1}
 -{\rm det} (  \hat  g ) = \frac{\lambda^2 \sigma}{A\sigma^2+2B\sigma+C} 
\end{equation}
 with coefficients
 \begin{equation}\label{coeffs}
\begin{split} 
&    A = (\lambda_{\max}^2-\lambda^2) (\lambda^2-\lambda_{\min}^2)\ ,
\\[1ex]
& B = \lambda^2(M_1+M_2) - \lambda_0^2(M_1-M_2)\ ,
\\[1ex]
&    C = -(M_1-M_2)^2 + \lambda^2J_1^2\ . 
\end{split}
\end{equation}
The critical tensions in these expressions are
 \bea\label{3cr}
  \lambda_{\rm min}  =  \big\vert  {1\over \ell_1} - {1\over \ell_2}  \big\vert \ , \quad
 \lambda_{\rm max} = {1\over \ell_1}+  {1\over \ell_2}\ , \quad
 \lambda_0  = \sqrt{\lambda_{\rm max}\lambda_{\rm min}}\ . 
\eea
For a static wall, {\it i.e.} when $J_1=J_2=0$,  
 the above formulae reduce, as they should,  to the ones obtained
in  ref.\cite{Bachas:2021fqo}.\,\footnote{\,When comparing with this reference  beware that it 
uses  the    (slightly  confusing) notation 
 $ \hat g_{\sigma\sigma} \equiv g(\sigma)$ so that,  since   the metric  is diagonal  in the static case,   det\,$\hat g = -\sigma g(\sigma)$.  
  }  
The only effect of the non-zero $J_j$  is actually  to shift  the  coefficient $C$ in \eqref{coeffs}.

 The roots of the quadratic polynomial in the denominator of \eqref{det1}, 
 \bea\label{spmapp} 
 \sigma_\pm =  {-B \pm \sqrt{B^2 - AC} \over A} \ , 
 \eea
 determine the   behaviour of the solution. 
 If $\sigma_+$ is   either complex or negative  (part of) the brane worldvolume  has  
${\rm det}\, \hat g >0$  in  the ergoregion, 
so it is  spacelike and physically unacceptable. 
Acceptable solutions  have  $\sigma_+ >0$ or
$\sigma_+=0$, and  describe walls that avoid, respectively enter the ergoregion 
as explained  in the main text, see  section \ref{sec:5n}.

    The actual shape of the wall is found by inserting \eqref{det1} in \eqref{det} and solving for  $x_j'^{\,2} $. 
After some rearrangements this gives
\begin{equation}
   \epsilon_1  \frac{x_1'}{\ell_1} =  \frac{(\lambda^2+\lambda_0^2)\sigma + (M_1-M_2)} {2(\sigma+M_1\ell_1^2+J^2\ell_1^2/4\sigma) 
    \sqrt{A\sigma (\sigma - \sigma_+)(\sigma - \sigma_-) }}\ , 
\end{equation}
\begin{equation}
\epsilon_2  \frac{x_2'}{\ell_2} =     \frac{(\lambda^2-\lambda_0^2)\sigma - (M_1-M_2)} {2(\sigma+M_2\ell_2^2+J^2\ell_2^2/4\sigma) 
    \sqrt{A\sigma (\sigma - \sigma_+)(\sigma - \sigma_-) }}\ , 
\end{equation}
where $\epsilon_j=\pm$ are signs. They  are   fixed by   the linear equation \eqref{Isr1}
with the result
\bea
 \epsilon_j(\sigma)  =  -{\sigma \over \vert\sigma\vert}\ . 
\eea
These signs 
agree with the known universal
solution  \cite{Bachas:2002nz,Bachas:2021fqo}
near the AdS boundary, at $\sigma \to \infty$, and they 
 ensure that walls entering the ergoregion have no kink. 
 Expressing  the denominators   in terms of the horizon locations  \eqref{sHpm} gives 
   the equations 
   \eqref{xj11} and \eqref{xj22} of the main text.

 It is worth noting that the tensionless ($\lambda\to 0$)  limit of our solution  is singular. Indeed, 
 on one hand  extremising   the brane action and ignoring its back-reaction 
  gives a  geodesic worldvolume,  but on the other hand for  $\lambda=0$ fluctuations of the string are unsuppressed.  
 In fact,  when  $\lambda$ is small the wall starts as a geodesic near the AdS boundary but   always departs
 significantly in the interior. In particular, a geodesic never enters the equilibrium horizon, whereas a tensile string can, 
 even if it is very light.

  
 \section{Horizon inequalities}\label{app:B}
 
In section \ref{sec:6n} we  asserted  that  BTZ  geometries  whose ergoregions can be  glued together by  a thin brane obey
the inequalities 
    \begin{equation}\label{orderapp} 
   \sigma_+^{{\rm H}1} >  \sigma_+^{{\rm H}2}    \quad {\rm if} \ \ M_1 > M_2\, ; 
\qquad 
  \,  \sigma_+^{{\rm H}2} <  \sigma_+^{{\rm H}1}    \quad {\rm if} \ \  M_1< M_2 \  , 
 \end{equation}  
where the horizon locations are  
\bea\label{sHapp}
\sigma_{\pm}^{{\rm H}j} =  - {M_j\ell_j^2\over 2}   \pm  {1 \over 2}\sqrt{M_j^2 \ell_j^4 -J^2 \ell_j^2 } \  
\eea  
and 
 $J  \equiv  \vert J_1\vert = \vert J_2\vert >0$. This ordering of the outer horizons is  manifest if one expands 
 at the  leading order for  small   $J$. 
We  want to show  that it is  valid  for all  values of   $J$. 

 If  as $J$ is cranked up  the ordering was  at some point   reversed,   then at this point we would   have
  $\sigma_+^{{\rm H}1} =   \sigma_+^{{\rm H}2} $, or equivalently  
\bea\label{b3}
 M_2\ell_2^2 - M_1\ell_1^2 =  \sqrt{ M_2^2 \ell_2^4 - J^2\ell_2^2} - \sqrt{ M_1^2 \ell_1^4 - J^2\ell_1^2} \ . 
\eea
Squaring twice   to eliminate the square roots gives
\begin{equation} \label{b4} 
    J^{\,2} = \frac{4\ell_1^2\ell_2^2(M_1-M_2)(M_2\ell_2^2-M_1\ell_1^2)}{(\ell_2^2-\ell_1^2)^2}  \ . 
\end{equation}
Without loss of generality we  assume, as elsewhere in the  text, 
 that $\ell_1 \leq \ell_2$. If $M_2>M_1$,  then automatically $M_2\ell_2^2 > M_1\ell_1^2$
and   \eqref{b4} has no 
solution for real $J$. In this case the ordering  \eqref{orderapp} cannot be reversed. 

 If on the other hand $M_1>M_2$ and $M_2\ell_2^2-M_1 \ell_1^2 >  0$  we need to  work harder. 
 Inserting $J^2$  from  (\ref{b4})  back in the original equation \eqref{b3} gives after   rearrangements
  \begin{equation}
\begin{aligned}
    (\ell_2^2-\ell_1^2) (M_2\ell_2^2-M_1 \ell_1^2) \, =\ \,  &\ell_2^2\, \bigl|
   (M_2\ell_2^2-M_1\ell_1^2) -  \ell_1^2(M_1-M_2)
    \bigr| \\
    &- \ell_1^2\left| (M_2\ell_2^2-M_1\ell_1^2) -\ell_2^2(M_1-M_2) \right| \ , 
\end{aligned}
\end{equation}
where the absolute values come from the square roots. This equation is not automatically obeyed whenever
its doubly-squared version is. 
 A solution  only exists
if 
\begin{equation}
    M_1-M_2\,\leq\, \frac{\ell_2^2M_2-\ell_1^2M_1}{\ell_2^2}
    \label{eq:existenceineq} \ \Leftrightarrow\  \frac{M_1}{M_2}\,\leq\, \frac{2\ell_2^2}{\ell_2^2+\ell_1^2} \ \  . 
\end{equation}
Remember now  that we only  care about solutions with walls in  the ergoregion, for which  
$M_1-M_2= \lam J$, see eq.\eqref{s+0}. Plugging  in \eqref{b4} this gives
\begin{equation}
     M_2 =   \bigl[ 1 - \frac{4\lam_0^2\, \lam^2  }{\lam_0^{\,4} +4 {\lam^2}/{\ell_1^2}} \bigr] M_1
    \label{eq:JsolT}
\end{equation}
with $\lambda_0^2 = (\ell_2^2 -   \ell_1^2)/\ell_1^2\ell_2^2$\,, see eq.\eqref{3crtens}. 
Consistency  with the bound   \eqref{eq:existenceineq} for a   brane  tension   in the  allowed range then 
requires
\begin{equation}
  \lambda_{\rm min} <   \lam\, \leq\,  \frac{\ell_1\lam_0^2}{2}\ , 
\end{equation}
where $ \lambda_{\rm min} = (\ell_2-\ell_1)/\ell_1\ell_2$. As one  can easily check,  this implies  $\ell_1 > \ell_2$ 
which  contradicits   our initial assumption. We conclude that  
 \eqref{b3} has no solutiion, and   the ordering
 \eqref{orderapp} holds for all $J$,   {\footnotesize QED}. 

\smallskip

   For completeness, let us also consider  the ordering of the inner horizons. 
Clearly $\sigma_+^{{\rm H}j} >    \sigma_-^{{\rm H}j}$ always, and for small $J$  also  
$\sigma_+^{{\rm H}1} >    \sigma_-^{{\rm H}2}$ and $\sigma_+^{{\rm H}2} >    \sigma_-^{{\rm H}1}$. 
To violate  these last inequalities we need  $ \sigma_+^{{\rm H}1} =   \sigma_-^{{\rm H}2}$ or  
 $ \sigma_+^{{\rm H}2} =   \sigma_-^{{\rm H}1}$
     for some finite $J$, or equivalently 
\bea\label{b9} 
 M_2\ell_2^2 - M_1\ell_1^2 =  \mp \bigl( \sqrt{ M_2^2 \ell_2^4 -  J^2\ell_2^2} + \sqrt{ M_1^2 \ell_1^4 - J^2\ell_1^2} \,\bigr)  \ . 
\eea
 Squaring twice gives back  eq.\eqref{b4} which has no  solution if   $M_2>M_1$.   
But if  $M_1>M_2$ and $M_2\ell_2^2-M_1 \ell_1^2 >  0$,  solutions to $ \sigma_+^{{\rm H}2} =   \sigma_-^{{\rm H}1}$
cannot   be ruled out. Indeed, inserting $J$ from \eqref{b4} in \eqref{b9} with the $+$ sign gives 
   \begin{equation}
\begin{aligned}
    (\ell_2^2-\ell_1^2) (M_2\ell_2^2-M_1 \ell_1^2) \, =\ \,  &\ell_2^2\, \bigl|
   (M_2\ell_2^2-M_1\ell_1^2) -  \ell_1^2(M_1-M_2)
    \bigr| \\
    & + \ell_1^2\left| (M_2\ell_2^2-M_1\ell_1^2) -\ell_2^2(M_1-M_2) \right| \ , 
\end{aligned}
\end{equation}
which requires that 
 \bea
  \frac{\ell_2^2M_2-\ell_1^2M_1}{\ell_2^2} \,\leq\,   M_1-M_2\,\leq\, \frac{\ell_2^2M_2-\ell_1^2M_1}{\ell_1^2}\ . 
 \eea
These conditions are  compatible with  $M_1-M_2= \lam J$ and $\lam$ in the allowed range, so
the outer horizon of slice 2 need not  always come before  the Cauchy horizon
of slice 1. 

   Finally one may  ask if the   inner (Cauchy)  horizons can join  continuously, i.e. if 
  $\sigma_-^{{\rm H}1} =  \sigma_-^{{\rm H}2}$ is allowed.  A simple calculation shows that this is indeed possible 
  for $\ell_2/\ell_1 <3$, a   critical ratio
  of central charges that also  arose  in  references \cite{Simidzija:2020ukv,Bachas:2021fqo}.
  We don't know if this is  a coincidence,  or if  some deeper reason lurks behind.

  
 \section{Background on flowing  funnels}\label{app:C}
 
   In this appendix we collect some  formulae 
   on the  flowing funnels discussed  in section \ref{sec:61n}. 
   We start with 
   the most general asymptotically-locally-AdS$_3$  solution in  Fefferman-Graham coordinates,  generalising  the
   Banados geometries \eqref{FG} to arbitrary boundary metric 
  \cite{Skenderis:1999nb}
      \bea\label{FGapp}
     ds^2 =   {\ell^2  dz^2 \over z^2} + {1\over z^2}\, g_{\alpha\beta}(x, z) \, dx^\alpha dx^\beta\ , 
     \eea
where $g_{\alpha\beta}$ is a quartic polynomial in $z$ (written here as a matrix) 
  \bea\label{FG1app}
   g(x,z)    
   =\ g_{(0)} + z^2 g_{(2)} +  {z^4\over 4} g_{(2)}\,g_{(0)}^{-1}\,g_{(2)}  \ . 
  \eea
In this equation  $g_{(0)}$ is the boundary metric and $g_{(2)}$ is  given by 
 \bea\label{FG2app}
  g_{(2)\,\alpha\beta}  = -  {\ell^2\over 2 }  R_{(0)}\, g_{(0)\,\alpha\beta} +  {\ell  }\,  \langle T_{\alpha\beta}\rangle  \ , 
  \eea
where  $R_{(0)}$ is the  Ricci scalar  of $g_{(0)}$,
and $\langle T_{\alpha\beta}\rangle$ the expectation value of the energy-momentum tensor. 
This 
  must   be  conserved, 
    $\nabla^a_{(0)}  \langle T_{ab}\rangle = 0$\,,
   and  should obey the  trace  equation 
    \,$g_{(0)}^{ab} \langle T_{ab}\rangle = (c /24\pi)    \, R_{(0)}$.
    
    \smallskip

 We  may take  the   boundary metric to be  that of the Schwarzschild  black hole
 (this differs from the metric in \cite{Fischetti:2012ps}, but since it is not dynamical we are free to  choose our preferred 
 boundary metric), 
 \bea
  ds^2_{(0)} =  -f(x)\, dt^2 + {dx^2\over f(x)} \qquad {\rm with} \quad f(x) = {x \over x + a }\ . 
 \eea
The horizon at $x= 0 $  has  temperature $\Theta_{S} = (4\pi a)^{-1}$.  Using the familiar tortoise coordinates we can write
 \bea
    ds^2_{(0)} =   f(x) (-dt^2 + dx_*^2)  \qquad {\rm where} \quad x_*  =  x + a \log x \ .
 \eea
 Let  $w^\pm = x_*  \pm t$.  
 The expectation value 
  of the energy-momentum tensor in the black-hole  metric can be  expressed in terms of   $ \phi = \log f(x)$
 as follows
\bea\label{C5} 
 \langle T_{\pm\pm}\rangle = {\ell\over 2}\, \big[ \partial^2_\pm\phi - {1\over 2} (\partial_\pm \phi)^2 \big]   + k_\pm(w^\pm)\ ,  \quad
   \langle T_{+-}\rangle =   -{\ell\over 2}\,  \partial_+\partial_-\phi\ , 
\eea
 with  $k_\pm$  arbitrary functions of $w^\pm$ that depend on the choice of state. At
$x\gg a$  where the metric is flat,    $k_\pm$ determine   the incoming and outgoing fluxes of energy. In   a stationary solution  these
 must be constant. If a heat bath 
 at temperature $\Theta_+$ is placed at infinity,   $k_+ = \pi^2\ell \, \Theta_+^2$\,. 
The function $k_-$,  on the other hand,  is fixed by  requiring that there is no outgoing flux at the Schwarzschild horizon.  
From 
\bea\label{C7}
{\ell\over 2}\, \big[ \partial^2_\pm\phi - {1\over 2} (\partial_\pm \phi)^2 \big] =  - {\ell \, (a^2 + 4ax)\over 16 (x+a)^4} \,  
\eea
we deduce   $ \langle T_{--}\rangle\vert_{x=0} = 0  \,\Longrightarrow\, k_- = \ell/16 a^2 = \pi^2\ell \, \Theta_{S}^2$\,. 
The outgoing flux at infinity  is thermalised at the black hole temperature,  as expected. 
 \smallskip
 
   Inserting the expressions  (\ref{FG2app}\,-\,\ref{C7})  in   eqs.\,\eqref{FGapp} and \eqref{FG1app} 
   gives the flowing-funnel metric in Fefferman-Graham  coordinates. These are however singular coordinates, not well
   adapted for calculating the event horizon as shown in \cite{Fischetti:2012ps}.  Following this reference, 
 one can compute  the horizon by going  to 
  BTZ coordinates -- this is possible because all solutions are  locally equivalent in three dimensions.
  The change   from any metric   \eqref{FGapp}\,-\,\eqref{FG1app}  to  local BTZ coordinates
  has been worked out in ref.\,\cite{Rooman:2000ei} (see also \cite{Krasnov:2001cu}) and can be used to compute the
  black-funnel shapes.  A noteworthy feature   is that the funnels  start  vertically inwards
  at $x=0$ \cite{Fischetti:2012ps} making  a delta-function contribution to the area density. 
  Note that figure \ref{fig:bfun} shows two independent flowing funnels with Schwarzschild  temperatures $\Theta_S = \Theta_1^{\rm eff}$
  and $\Theta_2^{\rm eff}$.

 
\bibliography{CVZ}{}

\providecommand{\href}[2]{#2}\begingroup\raggedright\begin{thebibliography}{10}

\bibitem{Maldacena:1997re}
J.~M. Maldacena, ``{The Large N limit of superconformal field theories and
  supergravity},'' \href{http://dx.doi.org/10.1023/A:1026654312961}{{\em Adv.
  Theor. Math. Phys.} {\bfseries 2} (1998) 231--252},
  \href{http://arxiv.org/abs/hep-th/9711200}{{\ttfamily arXiv:hep-th/9711200}}.

\bibitem{Gubser:1998bc}
S.~S. Gubser, I.~R. Klebanov, and A.~M. Polyakov, ``{Gauge theory correlators
  from noncritical string theory},''
  \href{http://dx.doi.org/10.1016/S0370-2693(98)00377-3}{{\em Phys. Lett. B}
  {\bfseries 428} (1998) 105--114},
  \href{http://arxiv.org/abs/hep-th/9802109}{{\ttfamily arXiv:hep-th/9802109}}.

\bibitem{Witten:1998qj}
E.~Witten, ``{Anti-de Sitter space and holography},''
  \href{http://dx.doi.org/10.4310/ATMP.1998.v2.n2.a2}{{\em Adv. Theor. Math.
  Phys.} {\bfseries 2} (1998) 253--291},
  \href{http://arxiv.org/abs/hep-th/9802150}{{\ttfamily arXiv:hep-th/9802150}}.

\bibitem{Liu:2018crr}
H.~Liu and J.~Sonner, ``{Holographic systems far from equilibrium: a review},''
  \href{http://arxiv.org/abs/1810.02367}{{\ttfamily arXiv:1810.02367
  [hep-th]}}.

\bibitem{Bernard:2016nci}
D.~Bernard and B.~Doyon, ``{Conformal field theory out of equilibrium: a
  review},'' \href{http://dx.doi.org/10.1088/1742-5468/2016/06/064005}{{\em J.
  Stat. Mech.} {\bfseries 1606} no.~6, (2016) 064005},
  \href{http://arxiv.org/abs/1603.07765}{{\ttfamily arXiv:1603.07765
  [cond-mat.stat-mech]}}.

\bibitem{Banados:1992wn}
M.~Banados, C.~Teitelboim, and J.~Zanelli, ``{The Black hole in
  three-dimensional space-time},''
  \href{http://dx.doi.org/10.1103/PhysRevLett.69.1849}{{\em Phys. Rev. Lett.}
  {\bfseries 69} (1992) 1849--1851},
  \href{http://arxiv.org/abs/hep-th/9204099}{{\ttfamily arXiv:hep-th/9204099}}.

\bibitem{Banados:1992gq}
M.~Banados, M.~Henneaux, C.~Teitelboim, and J.~Zanelli, ``{Geometry of the
  (2+1) black hole},'' \href{http://dx.doi.org/10.1103/PhysRevD.48.1506}{{\em
  Phys. Rev. D} {\bfseries 48} (1993) 1506--1525},
  \href{http://arxiv.org/abs/gr-qc/9302012}{{\ttfamily arXiv:gr-qc/9302012}}.
  [Erratum: Phys.Rev.D 88, 069902 (2013)].

\bibitem{Karch:2000gx}
A.~Karch and L.~Randall, ``{Open and closed string interpretation of SUSY CFT's
  on branes with boundaries},''
  \href{http://dx.doi.org/10.1088/1126-6708/2001/06/063}{{\em JHEP} {\bfseries
  06} (2001) 063}, \href{http://arxiv.org/abs/hep-th/0105132}{{\ttfamily
  arXiv:hep-th/0105132}}.

\bibitem{Bachas:2001vj}
C.~Bachas, J.~de~Boer, R.~Dijkgraaf, and H.~Ooguri, ``{Permeable conformal
  walls and holography},''
  \href{http://dx.doi.org/10.1088/1126-6708/2002/06/027}{{\em JHEP} {\bfseries
  06} (2002) 027}, \href{http://arxiv.org/abs/hep-th/0111210}{{\ttfamily
  arXiv:hep-th/0111210}}.

\bibitem{Bachas:2020yxv}
C.~Bachas, S.~Chapman, D.~Ge, and G.~Policastro, ``{Energy Reflection and
  Transmission at 2D Holographic Interfaces},''
  \href{http://dx.doi.org/10.1103/PhysRevLett.125.231602}{{\em Phys. Rev.
  Lett.} {\bfseries 125} no.~23, (2020) 231602},
  \href{http://arxiv.org/abs/2006.11333}{{\ttfamily arXiv:2006.11333
  [hep-th]}}.

\bibitem{Simidzija:2020ukv}
P.~Simidzija and M.~Van~Raamsdonk, ``{Holo-ween},''
  \href{http://dx.doi.org/10.1007/JHEP12(2020)028}{{\em JHEP} {\bfseries 12}
  (2020) 028}, \href{http://arxiv.org/abs/2006.13943}{{\ttfamily
  arXiv:2006.13943 [hep-th]}}.

\bibitem{Bachas:2021fqo}
C.~Bachas and V.~Papadopoulos, ``{Phases of Holographic Interfaces},''
  \href{http://dx.doi.org/10.1007/JHEP04(2021)262}{{\em JHEP} {\bfseries 04}
  (2021) 262}, \href{http://arxiv.org/abs/2101.12529}{{\ttfamily
  arXiv:2101.12529 [hep-th]}}.

\bibitem{Bhattacharyya:2009uu}
S.~Bhattacharyya and S.~Minwalla, ``{Weak Field Black Hole Formation in
  Asymptotically AdS Spacetimes},''
  \href{http://dx.doi.org/10.1088/1126-6708/2009/09/034}{{\em JHEP} {\bfseries
  09} (2009) 034}, \href{http://arxiv.org/abs/0904.0464}{{\ttfamily
  arXiv:0904.0464 [hep-th]}}.

\bibitem{Hubeny:2009ru}
V.~E. Hubeny, D.~Marolf, and M.~Rangamani, ``{Hawking radiation in large N
  strongly-coupled field theories},''
  \href{http://dx.doi.org/10.1088/0264-9381/27/9/095015}{{\em Class. Quant.
  Grav.} {\bfseries 27} (2010) 095015},
  \href{http://arxiv.org/abs/0908.2270}{{\ttfamily arXiv:0908.2270 [hep-th]}}.

\bibitem{Fischetti:2012ps}
S.~Fischetti and D.~Marolf, ``{Flowing Funnels: Heat sources for field theories
  and the AdS$_3$ dual of CFT$_2$ Hawking radiation},''
  \href{http://dx.doi.org/10.1088/0264-9381/29/10/105004}{{\em Class. Quant.
  Grav.} {\bfseries 29} (2012) 105004},
  \href{http://arxiv.org/abs/1202.5069}{{\ttfamily arXiv:1202.5069 [hep-th]}}.

\bibitem{Emparan:2013fha}
R.~Emparan and M.~Martinez, ``{Black String Flow},''
  \href{http://dx.doi.org/10.1007/JHEP09(2013)068}{{\em JHEP} {\bfseries 09}
  (2013) 068}, \href{http://arxiv.org/abs/1307.2276}{{\ttfamily arXiv:1307.2276
  [hep-th]}}.

\bibitem{Marolf:2013ioa}
D.~Marolf, M.~Rangamani, and T.~Wiseman, ``{Holographic thermal field theory on
  curved spacetimes},''
  \href{http://dx.doi.org/10.1088/0264-9381/31/6/063001}{{\em Class. Quant.
  Grav.} {\bfseries 31} (2014) 063001},
  \href{http://arxiv.org/abs/1312.0612}{{\ttfamily arXiv:1312.0612 [hep-th]}}.

\bibitem{Santos:2020kmq}
J.~E. Santos, ``{To go or not to go with the flow: Hawking radiation at strong
  coupling},'' \href{http://dx.doi.org/10.1007/JHEP06(2020)104}{{\em JHEP}
  {\bfseries 06} (2020) 104}, \href{http://arxiv.org/abs/2003.05454}{{\ttfamily
  arXiv:2003.05454 [hep-th]}}.

\bibitem{Hubeny:2011hd}
V.~E. Hubeny, S.~Minwalla, and M.~Rangamani, ``{The fluid/gravity
  correspondence},'' in {\em {Theoretical Advanced Study Institute in
  Elementary Particle Physics}: {String theory and its Applications: From meV
  to the Planck Scale}}.
\newblock 7, 2011.
\newblock \href{http://arxiv.org/abs/1107.5780}{{\ttfamily arXiv:1107.5780
  [hep-th]}}.

\bibitem{Hawking:1971vc}
S.~W. Hawking, ``{Black holes in general relativity},''
  \href{http://dx.doi.org/10.1007/BF01877517}{{\em Commun. Math. Phys.}
  {\bfseries 25} (1972) 152--166}.

\bibitem{Hollands:2006rj}
S.~Hollands, A.~Ishibashi, and R.~M. Wald, ``{A Higher dimensional stationary
  rotating black hole must be axisymmetric},''
  \href{http://dx.doi.org/10.1007/s00220-007-0216-4}{{\em Commun. Math. Phys.}
  {\bfseries 271} (2007) 699--722},
  \href{http://arxiv.org/abs/gr-qc/0605106}{{\ttfamily arXiv:gr-qc/0605106}}.

\bibitem{Moncrief:2008mr}
V.~Moncrief and J.~Isenberg, ``{Symmetries of Higher Dimensional Black
  Holes},'' \href{http://dx.doi.org/10.1088/0264-9381/25/19/195015}{{\em Class.
  Quant. Grav.} {\bfseries 25} (2008) 195015},
  \href{http://arxiv.org/abs/0805.1451}{{\ttfamily arXiv:0805.1451 [gr-qc]}}.

\bibitem{Hawking:1973uf}
S.~W. Hawking and G.~F.~R. Ellis,
  \href{http://dx.doi.org/10.1017/CBO9780511524646}{{\em {The Large Scale
  Structure of Space-Time}}}.
\newblock Cambridge Monographs on Mathematical Physics. Cambridge University
  Press, 2, 2011.

\bibitem{Bernard:2014qia}
D.~Bernard, B.~Doyon, and J.~Viti, ``{Non-Equilibrium Conformal Field Theories
  with Impurities},''
  \href{http://dx.doi.org/10.1088/1751-8113/48/5/05FT01}{{\em J. Phys. A}
  {\bfseries 48} no.~5, (2015) 05FT01},
  \href{http://arxiv.org/abs/1411.0470}{{\ttfamily arXiv:1411.0470 [math-ph]}}.

\bibitem{Meineri:2019ycm}
M.~Meineri, J.~Penedones, and A.~Rousset, ``{Colliders and conformal
  interfaces},'' \href{http://dx.doi.org/10.1007/JHEP02(2020)138}{{\em JHEP}
  {\bfseries 02} (2020) 138}, \href{http://arxiv.org/abs/1904.10974}{{\ttfamily
  arXiv:1904.10974 [hep-th]}}.

\bibitem{Carlip:1995qv}
S.~Carlip, ``{The (2+1)-Dimensional black hole},''
  \href{http://dx.doi.org/10.1088/0264-9381/12/12/005}{{\em Class. Quant.
  Grav.} {\bfseries 12} (1995) 2853--2880},
  \href{http://arxiv.org/abs/gr-qc/9506079}{{\ttfamily arXiv:gr-qc/9506079}}.

\bibitem{Chang:2013gba}
H.-C. Chang, A.~Karch, and A.~Yarom, ``{An ansatz for one dimensional steady
  state configurations},''
  \href{http://dx.doi.org/10.1088/1742-5468/2014/06/P06018}{{\em J. Stat.
  Mech.} {\bfseries 1406} no.~6, (2014) P06018},
  \href{http://arxiv.org/abs/1311.2590}{{\ttfamily arXiv:1311.2590 [hep-th]}}.

\bibitem{Bhaseen:2013ypa}
M.~J. Bhaseen, B.~Doyon, A.~Lucas, and K.~Schalm, ``{Far from equilibrium
  energy flow in quantum critical systems},''
  \href{http://dx.doi.org/10.1038/nphys3220}{{\em Nature Phys.} {\bfseries 11}
  (2015) 5}, \href{http://arxiv.org/abs/1311.3655}{{\ttfamily arXiv:1311.3655
  [hep-th]}}.

\bibitem{Pourhasan:2015bsa}
R.~Pourhasan, ``{Non-equilibrium steady state in the hydro regime},''
  \href{http://dx.doi.org/10.1007/JHEP02(2016)005}{{\em JHEP} {\bfseries 02}
  (2016) 005}, \href{http://arxiv.org/abs/1509.01162}{{\ttfamily
  arXiv:1509.01162 [cond-mat.stat-mech]}}.

\bibitem{Erdmenger:2017gdk}
J.~Erdmenger, D.~Fernandez, M.~Flory, E.~Megias, A.-K. Straub, and
  P.~Witkowski, ``{Time evolution of entanglement for holographic steady state
  formation},'' \href{http://dx.doi.org/10.1007/JHEP10(2017)034}{{\em JHEP}
  {\bfseries 10} (2017) 034}, \href{http://arxiv.org/abs/1705.04696}{{\ttfamily
  arXiv:1705.04696 [hep-th]}}.

\bibitem{Craps:2020ahu}
B.~Craps, M.~De~Clerck, P.~Hacker, K.~Nguyen, and C.~Rabideau, ``{Slow
  scrambling in extremal BTZ and microstate geometries},''
  \href{http://dx.doi.org/10.1007/JHEP03(2021)020}{{\em JHEP} {\bfseries 03}
  (2021) 020}, \href{http://arxiv.org/abs/2009.08518}{{\ttfamily
  arXiv:2009.08518 [hep-th]}}.

\bibitem{Banados:1998gg}
M.~Banados, ``{Three-dimensional quantum geometry and black holes},''
  \href{http://dx.doi.org/10.1063/1.59661}{{\em AIP Conf. Proc.} {\bfseries
  484} no.~1, (1999) 147--169},
  \href{http://arxiv.org/abs/hep-th/9901148}{{\ttfamily arXiv:hep-th/9901148}}.

\bibitem{Ryu:2006bv}
S.~Ryu and T.~Takayanagi, ``{Holographic derivation of entanglement entropy
  from AdS/CFT},'' \href{http://dx.doi.org/10.1103/PhysRevLett.96.181602}{{\em
  Phys. Rev. Lett.} {\bfseries 96} (2006) 181602},
  \href{http://arxiv.org/abs/hep-th/0603001}{{\ttfamily arXiv:hep-th/0603001}}.

\bibitem{Hubeny:2007xt}
V.~E. Hubeny, M.~Rangamani, and T.~Takayanagi, ``{A Covariant holographic
  entanglement entropy proposal},''
  \href{http://dx.doi.org/10.1088/1126-6708/2007/07/062}{{\em JHEP} {\bfseries
  07} (2007) 062}, \href{http://arxiv.org/abs/0705.0016}{{\ttfamily
  arXiv:0705.0016 [hep-th]}}.

\bibitem{Strominger:1997eq}
A.~Strominger, ``{Black hole entropy from near horizon microstates},''
  \href{http://dx.doi.org/10.1088/1126-6708/1998/02/009}{{\em JHEP} {\bfseries
  02} (1998) 009}, \href{http://arxiv.org/abs/hep-th/9712251}{{\ttfamily
  arXiv:hep-th/9712251}}.

\bibitem{Sonner:2017jcf}
J.~Sonner and B.~Withers, ``{Universal spatial structure of nonequilibrium
  steady states},''
  \href{http://dx.doi.org/10.1103/PhysRevLett.119.161603}{{\em Phys. Rev.
  Lett.} {\bfseries 119} no.~16, (2017) 161603},
  \href{http://arxiv.org/abs/1705.01950}{{\ttfamily arXiv:1705.01950
  [hep-th]}}.

\bibitem{Novak:2018pnv}
I.~Novak, J.~Sonner, and B.~Withers, ``{Overcoming obstacles in nonequilibrium
  holography},'' \href{http://dx.doi.org/10.1103/PhysRevD.98.086023}{{\em Phys.
  Rev. D} {\bfseries 98} no.~8, (2018) 086023},
  \href{http://arxiv.org/abs/1806.08655}{{\ttfamily arXiv:1806.08655
  [hep-th]}}.

\bibitem{Medenjak:2020bpe}
M.~Medenjak, G.~Policastro, and T.~Yoshimura, ``{Thermal transport in $T
  \bar{T}$-deformed conformal field theories: From integrability to
  holography},'' \href{http://dx.doi.org/10.1103/PhysRevD.103.066012}{{\em
  Phys. Rev. D} {\bfseries 103} no.~6, (2021) 066012},
  \href{http://arxiv.org/abs/2010.15813}{{\ttfamily arXiv:2010.15813
  [cond-mat.stat-mech]}}.

\bibitem{Medenjak:2020ppv}
M.~Medenjak, G.~Policastro, and T.~Yoshimura, ``{$T\bar{T}$-Deformed Conformal
  Field Theories out of Equilibrium},''
  \href{http://dx.doi.org/10.1103/PhysRevLett.126.121601}{{\em Phys. Rev.
  Lett.} {\bfseries 126} no.~12, (2021) 121601},
  \href{http://arxiv.org/abs/2011.05827}{{\ttfamily arXiv:2011.05827
  [cond-mat.stat-mech]}}.

\bibitem{Billo:2016cpy}
M.~Bill\`o, V.~Gon\c{c}alves, E.~Lauria, and M.~Meineri, ``{Defects in
  conformal field theory},''
  \href{http://dx.doi.org/10.1007/JHEP04(2016)091}{{\em JHEP} {\bfseries 04}
  (2016) 091}, \href{http://arxiv.org/abs/1601.02883}{{\ttfamily
  arXiv:1601.02883 [hep-th]}}.

\bibitem{Azeyanagi:2007qj}
T.~Azeyanagi, A.~Karch, T.~Takayanagi, and E.~G. Thompson, ``{Holographic
  calculation of boundary entropy},''
  \href{http://dx.doi.org/10.1088/1126-6708/2008/03/054}{{\em JHEP} {\bfseries
  03} (2008) 054}, \href{http://arxiv.org/abs/0712.1850}{{\ttfamily
  arXiv:0712.1850 [hep-th]}}.

\bibitem{Bachas:2007td}
C.~Bachas and I.~Brunner, ``{Fusion of conformal interfaces},''
  \href{http://dx.doi.org/10.1088/1126-6708/2008/02/085}{{\em JHEP} {\bfseries
  02} (2008) 085}, \href{http://arxiv.org/abs/0712.0076}{{\ttfamily
  arXiv:0712.0076 [hep-th]}}.

\bibitem{Bak:2011ga}
D.~Bak, M.~Gutperle, and R.~A. Janik, ``{Janus Black Holes},''
  \href{http://dx.doi.org/10.1007/JHEP10(2011)056}{{\em JHEP} {\bfseries 10}
  (2011) 056}, \href{http://arxiv.org/abs/1109.2736}{{\ttfamily arXiv:1109.2736
  [hep-th]}}.

\bibitem{penrose1974gravitational}
R.~Penrose, ``Gravitational collapse,'' in {\em Symposium-International
  Astronomical Union}, vol.~64, pp.~82--91, Cambridge University Press.
\newblock 1974.

\bibitem{Dias:2019ery}
O.~J.~C. Dias, H.~S. Reall, and J.~E. Santos, ``{The BTZ black hole violates
  strong cosmic censorship},''
  \href{http://dx.doi.org/10.1007/JHEP12(2019)097}{{\em JHEP} {\bfseries 12}
  (2019) 097}, \href{http://arxiv.org/abs/1906.08265}{{\ttfamily
  arXiv:1906.08265 [hep-th]}}.

\bibitem{Papadodimas:2019msp}
K.~Papadodimas, S.~Raju, and P.~Shrivastava, ``{A simple quantum test for
  smooth horizons},'' \href{http://dx.doi.org/10.1007/JHEP12(2020)003}{{\em
  JHEP} {\bfseries 12} (2020) 003},
  \href{http://arxiv.org/abs/1910.02992}{{\ttfamily arXiv:1910.02992
  [hep-th]}}.

\bibitem{Balasubramanian:2019qwk}
V.~Balasubramanian, A.~Kar, and G.~S\'arosi, ``{Holographic Probes of Inner
  Horizons},'' \href{http://dx.doi.org/10.1007/JHEP06(2020)054}{{\em JHEP}
  {\bfseries 06} (2020) 054}, \href{http://arxiv.org/abs/1911.12413}{{\ttfamily
  arXiv:1911.12413 [hep-th]}}.

\bibitem{Emparan:2020rnp}
R.~Emparan and M.~Toma\v{s}evi\'c, ``{Strong cosmic censorship in the BTZ black
  hole},'' \href{http://dx.doi.org/10.1007/JHEP06(2020)038}{{\em JHEP}
  {\bfseries 06} (2020) 038}, \href{http://arxiv.org/abs/2002.02083}{{\ttfamily
  arXiv:2002.02083 [hep-th]}}.

\bibitem{Pandya:2020ejc}
A.~Pandya and F.~Pretorius, ``{The rotating black hole interior: Insights from
  gravitational collapse in $AdS_3$ spacetime},''
  \href{http://dx.doi.org/10.1103/PhysRevD.101.104026}{{\em Phys. Rev. D}
  {\bfseries 101} no.~10, (2020) 104026},
  \href{http://arxiv.org/abs/2002.07130}{{\ttfamily arXiv:2002.07130 [gr-qc]}}.

\bibitem{Gibbons:2000xe}
G.~W. Gibbons and C.~A.~R. Herdeiro, ``{Born-Infeld theory and stringy
  causality},'' \href{http://dx.doi.org/10.1103/PhysRevD.63.064006}{{\em Phys.
  Rev. D} {\bfseries 63} (2001) 064006},
  \href{http://arxiv.org/abs/hep-th/0008052}{{\ttfamily arXiv:hep-th/0008052}}.

\bibitem{Frolov:1995qp}
V.~P. Frolov and A.~L. Larsen, ``{Stationary strings and 2-d black holes},''
  \href{http://dx.doi.org/10.1016/0550-3213(95)00302-9}{{\em Nucl. Phys. B}
  {\bfseries 449} (1995) 149--158},
  \href{http://arxiv.org/abs/hep-th/9503060}{{\ttfamily arXiv:hep-th/9503060}}.

\bibitem{Hawking:1982dh}
S.~W. Hawking and D.~N. Page, ``{Thermodynamics of Black Holes in anti-De
  Sitter Space},'' \href{http://dx.doi.org/10.1007/BF01208266}{{\em Commun.
  Math. Phys.} {\bfseries 87} (1983) 577}.

\bibitem{Witten:1998zw}
E.~Witten, ``{Anti-de Sitter space, thermal phase transition, and confinement
  in gauge theories},''
  \href{http://dx.doi.org/10.4310/ATMP.1998.v2.n3.a3}{{\em Adv. Theor. Math.
  Phys.} {\bfseries 2} (1998) 505--532},
  \href{http://arxiv.org/abs/hep-th/9803131}{{\ttfamily arXiv:hep-th/9803131}}.

\bibitem{Hubeny:2012ry}
V.~E. Hubeny, ``{Extremal surfaces as bulk probes in AdS/CFT},''
  \href{http://dx.doi.org/10.1007/JHEP07(2012)093}{{\em JHEP} {\bfseries 07}
  (2012) 093}, \href{http://arxiv.org/abs/1203.1044}{{\ttfamily arXiv:1203.1044
  [hep-th]}}.

\bibitem{AbajoArrastia:2010yt}
J.~Abajo-Arrastia, J.~Aparicio, and E.~Lopez, ``{Holographic Evolution of
  Entanglement Entropy},''
  \href{http://dx.doi.org/10.1007/JHEP11(2010)149}{{\em JHEP} {\bfseries 11}
  (2010) 149}, \href{http://arxiv.org/abs/1006.4090}{{\ttfamily arXiv:1006.4090
  [hep-th]}}.

\bibitem{Balasubramanian:2010ce}
V.~Balasubramanian, A.~Bernamonti, J.~de~Boer, N.~Copland, B.~Craps,
  E.~Keski-Vakkuri, B.~Muller, A.~Schafer, M.~Shigemori, and W.~Staessens,
  ``{Thermalization of Strongly Coupled Field Theories},''
  \href{http://dx.doi.org/10.1103/PhysRevLett.106.191601}{{\em Phys. Rev.
  Lett.} {\bfseries 106} (2011) 191601},
  \href{http://arxiv.org/abs/1012.4753}{{\ttfamily arXiv:1012.4753 [hep-th]}}.

\bibitem{Hartman:2013qma}
T.~Hartman and J.~Maldacena, ``{Time Evolution of Entanglement Entropy from
  Black Hole Interiors},''
  \href{http://dx.doi.org/10.1007/JHEP05(2013)014}{{\em JHEP} {\bfseries 05}
  (2013) 014}, \href{http://arxiv.org/abs/1303.1080}{{\ttfamily arXiv:1303.1080
  [hep-th]}}.

\bibitem{Bhattacharyya:2008xc}
S.~Bhattacharyya, V.~E. Hubeny, R.~Loganayagam, G.~Mandal, S.~Minwalla,
  T.~Morita, M.~Rangamani, and H.~S. Reall, ``{Local Fluid Dynamical Entropy
  from Gravity},'' \href{http://dx.doi.org/10.1088/1126-6708/2008/06/055}{{\em
  JHEP} {\bfseries 06} (2008) 055},
  \href{http://arxiv.org/abs/0803.2526}{{\ttfamily arXiv:0803.2526 [hep-th]}}.

\bibitem{Engelhardt:2014gca}
N.~Engelhardt and A.~C. Wall, ``{Quantum Extremal Surfaces: Holographic
  Entanglement Entropy beyond the Classical Regime},''
  \href{http://dx.doi.org/10.1007/JHEP01(2015)073}{{\em JHEP} {\bfseries 01}
  (2015) 073}, \href{http://arxiv.org/abs/1408.3203}{{\ttfamily arXiv:1408.3203
  [hep-th]}}.

\bibitem{Faulkner:2013ana}
T.~Faulkner, A.~Lewkowycz, and J.~Maldacena, ``{Quantum corrections to
  holographic entanglement entropy},''
  \href{http://dx.doi.org/10.1007/JHEP11(2013)074}{{\em JHEP} {\bfseries 11}
  (2013) 074}, \href{http://arxiv.org/abs/1307.2892}{{\ttfamily arXiv:1307.2892
  [hep-th]}}.

\bibitem{Bachas:2012bj}
C.~Bachas, I.~Brunner, and D.~Roggenkamp, ``{A worldsheet extension of
  O(d,d:Z)},'' \href{http://dx.doi.org/10.1007/JHEP10(2012)039}{{\em JHEP}
  {\bfseries 10} (2012) 039}, \href{http://arxiv.org/abs/1205.4647}{{\ttfamily
  arXiv:1205.4647 [hep-th]}}.

\bibitem{Lucas:2015hnv}
A.~Lucas, K.~Schalm, B.~Doyon, and M.~J. Bhaseen, ``{Shock waves, rarefaction
  waves, and nonequilibrium steady states in quantum critical systems},''
  \href{http://dx.doi.org/10.1103/PhysRevD.94.025004}{{\em Phys. Rev. D}
  {\bfseries 94} no.~2, (2016) 025004},
  \href{http://arxiv.org/abs/1512.09037}{{\ttfamily arXiv:1512.09037
  [hep-th]}}.

\bibitem{Spillane:2015daa}
M.~Spillane and C.~P. Herzog, ``{Relativistic Hydrodynamics and Non-Equilibrium
  Steady States},''
  \href{http://dx.doi.org/10.1088/1742-5468/2016/10/103208}{{\em J. Stat.
  Mech.} {\bfseries 1610} no.~10, (2016) 103208},
  \href{http://arxiv.org/abs/1512.09071}{{\ttfamily arXiv:1512.09071
  [hep-th]}}.

\bibitem{Ecker:2021ukv}
C.~Ecker, J.~Erdmenger, and W.~Van Der~Schee, ``{Non-equilibrium steady state
  formation in 3+1 dimensions},''
  \href{http://arxiv.org/abs/2103.10435}{{\ttfamily arXiv:2103.10435
  [hep-th]}}.

\bibitem{Lanczos}
K.~Lanczos, ``{Fl\"achenhafte Verteilung der Materie in der Einsteinschen
  Gravitationstheorie},'' {\em Annalen der Physik} {\bfseries 74} (1924) 518.

\bibitem{Israel:1966rt}
W.~Israel, ``{Singular hypersurfaces and thin shells in general relativity},''
  \href{http://dx.doi.org/10.1007/BF02710419}{{\em Nuovo Cim. B} {\bfseries
  44S10} (1966) 1}. [Erratum: Nuovo Cim.B 48, 463 (1967)].

\bibitem{Papapetrou}
A.~Papapetrou and A.~Hamoui, ``{Couches simples de mati\`ere en relativit\'e
  g\'en\'erale},'' {\em Ann.\,Inst.\,Henri Poincar\'e} {\bfseries IX} (1968)
  179--211.

\bibitem{Bachas:2002nz}
C.~Bachas, ``{Asymptotic symmetries of AdS$_2$ branes},'' in {\em {Meeting on
  Strings and Gravity: Tying the Forces Together}}.
\newblock 2001.
\newblock \href{http://arxiv.org/abs/hep-th/0205115}{{\ttfamily
  arXiv:hep-th/0205115}}.

\bibitem{Skenderis:1999nb}
K.~Skenderis and S.~N. Solodukhin, ``{Quantum effective action from the AdS /
  CFT correspondence},''
  \href{http://dx.doi.org/10.1016/S0370-2693(99)01467-7}{{\em Phys. Lett. B}
  {\bfseries 472} (2000) 316--322},
  \href{http://arxiv.org/abs/hep-th/9910023}{{\ttfamily arXiv:hep-th/9910023}}.

\bibitem{Rooman:2000ei}
M.~Rooman and P.~Spindel, ``{Uniqueness of the asymptotic AdS(3) geometry},''
  \href{http://dx.doi.org/10.1088/0264-9381/18/11/309}{{\em Class. Quant.
  Grav.} {\bfseries 18} (2001) 2117--2124},
  \href{http://arxiv.org/abs/gr-qc/0011005}{{\ttfamily arXiv:gr-qc/0011005}}.

\bibitem{Krasnov:2001cu}
K.~Krasnov, ``{On holomorphic factorization in asymptotically AdS 3-D
  gravity},'' \href{http://dx.doi.org/10.1088/0264-9381/20/18/311}{{\em Class.
  Quant. Grav.} {\bfseries 20} (2003) 4015--4042},
  \href{http://arxiv.org/abs/hep-th/0109198}{{\ttfamily arXiv:hep-th/0109198}}.

\end{thebibliography}\endgroup

\end{document}